 \documentclass[preprint, 5p, times, twocolumn]{elsarticle}

\usepackage{amssymb}
\usepackage{amsmath}
\usepackage{graphicx}
\usepackage[labelformat=simple]{subcaption}

\usepackage{enumerate}
\usepackage{stfloats}
\usepackage{nameref}
\usepackage{hyperref}
\hypersetup{colorlinks,linkcolor={blue},citecolor={blue},urlcolor={blue}} 
\usepackage[linesnumbered,ruled, vlined]{algorithm2e}
\usepackage{algpseudocode}
\usepackage{array}
\usepackage{textcomp}
\usepackage{stfloats}
\usepackage{url}
\usepackage{multirow}
\usepackage{tabularray}
\usepackage{booktabs}
\usepackage{adjustbox} 
\usepackage{colortbl}
\usepackage{nameref}
\usepackage{orcidlink}
\biboptions{sort&compress}

\usepackage{titlesec}
\titleformat{\paragraph}
{\normalfont\normalsize\itshape}{\theparagraph}{1em}{}
\titlespacing*{\paragraph}
{0pt}{3.25ex plus 1ex minus .2ex}{1.5ex plus .2ex}

\usepackage{copyrightbox}
\journal{Future Generation Computer Systems}

\makeatletter
\def\ps@pprintTitle{%
	\let\@oddhead\@empty
	\let\@evenhead\@empty
	\def\@oddfoot{\footnotesize\itshape
		\begin{minipage}{\textwidth}
			\centering
			0167-739X/© 2026 Elsevier B.V. All rights are reserved, including those for text and data mining, AI training, and similar technologies. \\
			The final published version is available at:
			\url{https://doi.org/10.1016/j.future.2026.108685}
	\end{minipage}}%
	\let\@evenfoot\@oddfoot}
\makeatother

\begin{document}
\begin{frontmatter}


\title{ReSAFT: An Efficient Stuck-at Fault-Tolerant Scheme for ReRAM-based Process-in-Memory Accelerators}

\author{Aniseh Dorostkar} 
\ead{a_dorostkar@aut.ac.ir}

\author{Hamed Farbeh\corref{cor1}} 
\ead{farbeh@aut.ac.ir}

\author{Hamid R. Zarandi}
\ead{h_zarandi@aut.ac.ir}

\cortext[cor1]{Corresponding author}

\address{Department of Computer Engineering, Amirkabir University of Technology (Tehran Polytechnic)}

\begin{abstract}	
Analog ReRAM-based process-in-memory (PIM) accelerators provide high parallelism and energy efficiency for deep convolutional neural networks (CNNs) inference. However, their susceptibility to permanent faults, such as \textit{stuck-at high} (\textit{SaH}) and \textit{stuck-at low} (\textit{SaL}) resistance states, poses a major challenge by permanently corrupting the CNN weights mapped to conductance values of ReRAM cells and degrading inference accuracy, which leads to system unreliability in safety-critical applications. In this paper, we propose a fault-tolerant scheme for analog ReRAM-based PIM accelerators to tackle stuck-at faults (SAFs) with minimal redundancy overhead to recover classification accuracy degradation. The proposed scheme contains a redundancy-based hardware solution alongside fault-aware mapping method for ensuring reliable analog computation in ReRAM crossbar. We analyze the impact of varying number of redundant rows and columns on accuracy and design metrics. Subsequently, a multi-objective optimization (MOO) problem is formulated and solved to efficiently determine the number of redundant rows and columns, considering trade-offs among various design metrics. Furthermore, a fault-aware weight mapping is proposed for dual-crossbar structures to further compensate for the accuracy degradation caused by SAFs. Simulation results show that, for the \textit{SimpleNet} model using the \textit{MNIST} dataset, the inference accuracy is recovered by approximately 22.39\%, on average, across four configurations of optimal solutions, each offering a trade-off between reliability and area, energy consumption, and latency overheads. The mean-time-to-failure (MTTF) improves by about $61\times$ on average compared to the baseline. These selected configurations also reduce energy and area overheads by 32\%, on average, in comparison to row-only and column-only configurations.
\end{abstract}


\end{frontmatter}


\section{Introduction}
\label{secIntro}
Deep neural networks (DNNs), particularly convolutional neural networks (CNNs), have greatly impacted applications such as healthcare, autonomous vehicles, space missions, and other safety-critical systems~\cite{Ibrahim2020}. Their remarkable performance in classification, detection, and recognition comes from deep, multi-layered architectures that extract features through vector-matrix multiplication (VMM) operations. However, increasing network complexity to achieve higher accuracy also significantly raises computational cost, memory/storage requirements, and energy consumption.

To meet the demanding computational needs of CNNs, conventional computing platforms based on CPUs, graph processing units (GPUs) and field-programmable gate arrays (FPGAs) are increasingly facing performance bottlenecks such as the memory wall and the power wall. Due to these limitations, processing-in-memory (PIM) paradigm have gained attention, which integrate computation directly within memory arrays, thereby reducing data movement and energy consumption. Among various memory technologies enabling PIM, Resistive Random-Access Memory (ReRAM) has emerged as a promising candidate due to its non-volatility, high density, low power consumption, and capability to perform analog PIM. Using the ReRAM crossbar structure, ReRAM-based PIM accelerators such as \textit{ISAAC}~\cite{Shafiee2016}, \textit{PRIME}~\cite{Chi2016}, and \textit{PipeLayer}~\cite{Song2017} can perform massively memory-intensive VMM operations in parallel, making them highly efficient for CNN inference.

Despite the mentioned advantages, due to the limitation of fabrication technology, ReRAM-based PIM accelerators suffer from process variations (PVs)~\cite{Liu2015, Grossi2019}, limited write endurance~\cite{Grossi2019, Beckmann2016}, sources of noises~\cite{He2019}, and nonlinearity of current–voltage characteristics~\cite{Chakraborty2020} and hardware faults~\cite{Chakraborty2020, Lammie2020, Li2019, Charan2020}. ReRAM faults are categorized into transient and permanent types based on their temporal characteristics~\cite{Gambardella2019}, and arise either in fabrication phase or operational phase. Transient faults are temporary and non-destructive which typically caused by disturbances like high energy particle hitting the electrical device, resistance drift, PVs, temperature fluctuations, or read/write disturbance~\cite{Ibrahim2020, Xia2019, Cheshmikhani2020}. These faults change the resistance of the cells without physically damage the ReRAM cells as they can still function, because the resistance of the cell can change.

In contrast, permanent faults are persistent over time and the resistance of cell is unchangeable. These faults are due to irreversible physical damage, often introduced by fabrication limitations (oxide layer thickness, process fluctuations, or reduced feature sizes~\cite{Yeo2019, Ching-YiChen2015}) or limitation in endurance~\cite{Beckmann2016,He2019}. Defects such as short defects and overforming stage defects~\cite{Ching-YiChen2015}, cause the resistance of a cell to stuck-at a resistance value (\textit{stuck-at high} (\textit{SaH}), \textit{stuck-at low} (\textit{SaL}) or \textit{stuck-at value} (\textit{SaV}) resistant state). \textit{SaH} faults are commonly caused by short defects that force the cell to remain in high resistance state (HRS), while \textit{SaL} faults may result from permanent cell failures, broken selectors, or damaged interconnects that force the cell to remain in low resistance state (LRS). Regardless of passing the manufacturing tests of ReRAM cells, permanent faults can occur over lifetime due to limited write endurance, excessive write-0 (or write-1) pulses~\cite{Grossi2019, Mittal2018}, or by damages to the device or transistor (e.g., due to wear out~\cite{Putra2023}).

Since NN parameters are mapped to the conductance values of ReRAM cells, any faults in ReRAM cells can easily corrupt these learned values, directly impact the outcomes of VMM, leading to computational inaccuracies. Consequently, process variations (PVs) and faults in ReRAM technology not only lower the yield of memory arrays but also impair the inference accuracy of CNNs executed on PIM accelerators~\cite{Song2021}. The vulnerability of ReRAM-based PIM accelerators in executing CNNs is highly sensitive to factors such as the type and depth of network layers, the location of learning parameters within each layer, as well as the value and types of the faults~\cite{Dorostkar2024}.

To address the reliability limitations of ReRAM-based PIM systems against faults, various software- and hardware-based fault-tolerant methods including mapping-based methods~\cite{Xia2019, Liu2017, Shin2023, Ma2023, Huang2022, Tung2023}, fault-aware training methods~\cite{Liu2015, Xia2019, Yeo2019, Wang2022, Dhingra2024}, error detection and correction codes~\cite{Liu2018, Lou2020, Wang2015, Peng2021, Pinto2024, Liu2020}, and redundancy-based methods~\cite{Xia2018}, have been proposed. Although software-based methods are cost-effective, most of them are not applicable to permanent faults due to the unchangeable resistance of faulty cells or become inefficient under high fault rates.

Conversely, existing hardware redundancy solutions, primarily designed for digital memory-centric applications, are not directly applicable to analog ReRAM-based PIM systems. In PIM crossbars, the fundamental computational unit is a column performing VMM; consequently, a single faulty cell can compromise the functionality of the entire column. Although 1T1R structures allow faulty cells to be isolated by controlling their associated access transistors~\cite{Chen2017}. this approach introduces substantial routing overhead. Furthermore, conventional redundancy-based methods typically rely solely on redundant rows or columns and deactivate an entire row or column once a fault is detected. However, these methods are often too coarse-grained for analog computing, resulting in substantial inefficiencies in area, power consumption, and latency that diminish the inherent energy and throughput advantages of PIM accelerators.

Consequently, there is a critical need for a solution that balances the trade-off between hardware cost, reliability, and inference accuracy in CNN models through integrated hardware- software-based solutions. Specifically, leveraging an optimal level of hardware and software redundancy mechanisms can effectively compensate for accuracy loss caused by SAFs.

This work is motivated by trading off the reliability, cost, and inference accuracy in analog ReRAM-based PIM accelerators. Our goal is to design a fault-tolerant architecture that tackles SAFs with minimal redundancy overhead while preserving CNN classification accuracy. To this end, we propose ReSAFT, a fault-tolerant scheme for analog ReRAM-based PIM accelerators that combines a redundancy-based ReRAM crossbar architecture with fault-aware mapping and remapping methods to tolerate SAF efficiently. The proposed fault-tolerant crossbar architecture integrates a combination of redundant rows and columns into the ReRAM crossbar structure with significantly fewer redundant resources compared to conventional redundancy schemes. We evaluate the impact of varying the number of redundant rows and columns on CNNs classification accuracy and design metrics, including reliability, energy consumption, area overhead, and latency. We then formulated and solved an architecture-level optimization problem to determine the adequate number of redundant rows and columns, a problem that requires solving complex multi-dimensional trade-offs among these design metrics. To efficiently utilize resources, we further propose software-level mapping and input scheduling across redundant rows and columns, and a dual-crossbar remapping method, which are applied at runtime. The main contributions of this article are as follows:
 
\begin{enumerate}
\item{
	We propose a hardware redundancy-based fault-tolerant crossbar architecture for analog ReRAM-based PIM accelerators. Unlike conventional methods that rely solely on either row or column redundancy, the proposed architecture employs a combination of redundant rows and columns that uses the strengths of both to tolerate \textit{SaH} and \textit{SaL} faults Specifically, redundant columns are used to tolerate both \textit{SaL} and \textit{SaH} faults, while redundant rows tolerate \textit{SaH} faults via current adjustment, significantly reducing area overhead compared to column-only methods.

}
\item{
	A multi-objective optimization (MOO) problem is formulated and solved to determine the adequate number of redundant rows and columns in ReRAM-based crossbars. The formulation considers trade-offs among design metrics including MTTF, energy consumption, latency, and area while satisfying a classification accuracy constraint.
}
\item{
	To efficiently use the hardware redundancy, a fault-aware mapping algorithm is proposed to sufficiently map faulty cells to appropriate redundant cells based on the specific fault type and location. This is an algorithmic enhancement to the PIM controller that works with the proposed redundant-based crossbar.
	}
\item{
	A complementary and low-cost fault-aware weight remapping is applied at runtime across dual ReRAM crossbars (contain positive and negative crossbars) for the remaining faults that cannot be tolerated by ReFTC architecture. This method ensures reliable computation while minimizing overhead and compensates for the accuracy degradation caused by SAFs.																																		

	} 
\end{enumerate}

The rest of this article is organized as follows. Section~\ref{preliminary} provides the background on NNs, ReRAM devices, and accelerator and a comprehensive review of related work. Section~\ref{resaft} describes the process of the proposed fault-tolerant scheme and details the redundancy-based crossbar design along with the proposed mapping and scheduling methods and evaluates the impact of varying the number of redundant rows and columns on design parameters. The simulation results and analysis of our methos are provided in section~\ref{evaluation}. Finally, section~\ref{Concl} concludes the article.

\section{Preliminaries and Related Work}
\label{preliminary}

\subsection{Convolutional Neural Networks}
CNNs are typically composed of a sequence of convolutional (\textit{CONV}) layers, pooling (\textit{POOL}) layers, fully-connected (\textit{FC}) layers, and optionally, custom layers such as Batch Normalization (\textit{BatchNorm})~\cite{Shafiee2016}. Trainable parameters, such as weights, kernels, and biases, are mainly in the \textit{CONV}, \textit{FC}, and \textit{BatchNorm} layers, and are adjusted during training phase. The main computations in CNNs involve multi-dimensional convolution and VMM. In particular, \textit{FC} layers are based on the VMM operation, which is performed at each neuron and represented by:
\begin{equation}
	\label{Eq1}
	y_j = \sigma\left( \sum_{i=0}^{n-1} x_i w_{ij} + b_j \right)
\end{equation}
where $y_{j}$  denotes the output of the \textit{j-th} neuron, $x_{i}$ is the input of the neuron (the output from \textit{i-th} neuron in the preceding layer), $w_{ij}$ refers to the synaptic weight associated with each input, and $b_j$ is the bias applied to each neuron. The weighted sum of inputs and bias is passed through an activation function $\sigma$, which can be \textit{sigmoid}, \textit{ReLU}, or \textit{tanh} function to compute the final neuron output. Additionally, in the \textit{CONV} layers, a convolution operation is applied between each kernel and its corresponding input feature map (\textit{IFM}) to generate the output feature map (\textit{OFM}), as defined by:
\begin{multline}
	\label{Eq2}
	\text{OFM}_n(x, y) =\sigma\sum_{i=0}^{C_{in}}\bigg(\sum_{dx=0}^{K_{W}}\bigg(\sum_{dy=0}^{K_{H}}	IFM(i,x+dx,y+dy){*}\\
	W_{n}(i,dx,dy)\bigg)+B_{i}\bigg)\ ,1\leq n \leq C_{out}
\end{multline}
where $IFM$ is the input feature map, $C_{in}$ is the number of input channels (feature maps), $C_{out}$ is the number of output channels, $W$ denotes the kernel matrix, $W_{n}(i,dx,dy)$ specifies \textit{n-th} kernel weights applied to the \textit{i-th} input channel, and $B_{i}$ is the bias array associated with each kernel. 

\subsection{ReRAM-based PIM Accelerators}
\begin{figure}[t]
	\centering
	\subfloat[\label{Fig1-a}]{
		\includegraphics[width=0.55\linewidth]{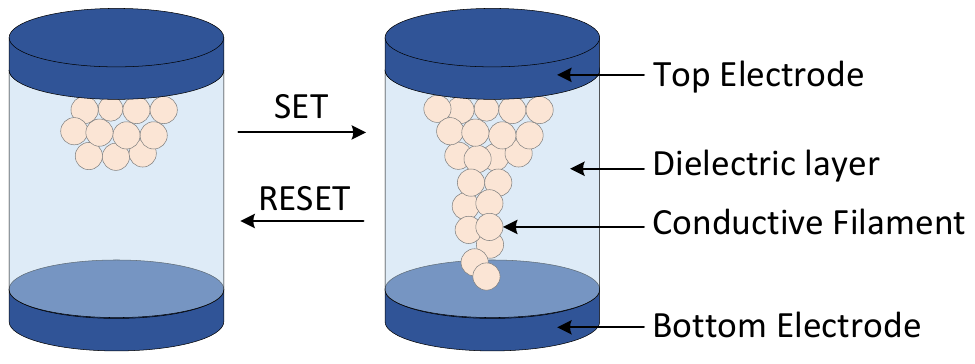}
	}
	\subfloat[\label{Fig1-b}]{
		\includegraphics[width=0.45\linewidth]{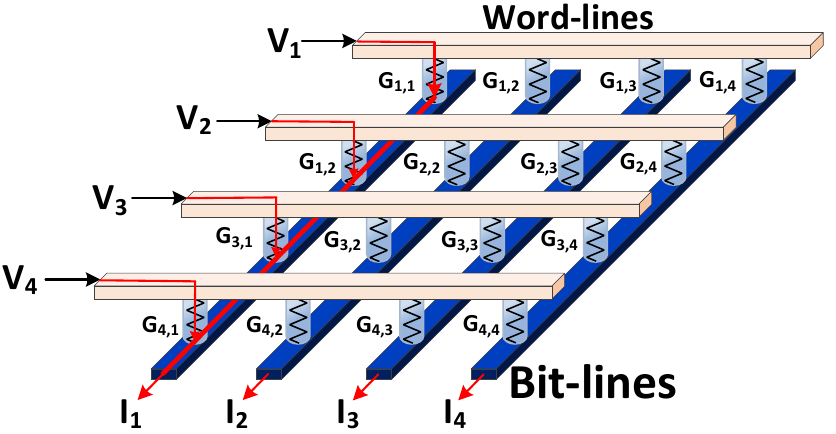}
	}
	\caption{ReRAM structure. a) a single ReRAM cell and b) ReRAM crossbar structure performing analog VMM operation.}
	\label{Fig1:multi2}
\end{figure}

ReRAM is a type of nonvolatile memory technology that stores data by utilizing variable resistance states in memory cells based on a metal–insulator–metal (MIM) structure. The resistance of each cell can be modulated by applying external voltages, allowing it to switch between different states. These resistance changes are controlled by the formation and rupture of conductive filaments within the insulating layer, as shown in Fig.~\ref{Fig1-a}. To read the stored data, a small read voltage is applied that is low enough to avoid disturbing the cell’s current state.

By arranging ReRAM cells in a crossbar structure, the system can naturally perform VMM in the analog domain. In this configuration, matrix elements are encoded as conductance values within the crossbar and input vectors are applied as voltage signals across the word-lines. According to \textit{Kirchhoff’s law}, the resulting currents collected on the bit-lines represent the output of the VMM operation, enabling efficient parallel computation in-memory (Fig.~\ref{Fig1-b}).

\begin{figure}[t]
	\centering
	\includegraphics[width=0.4\textwidth]{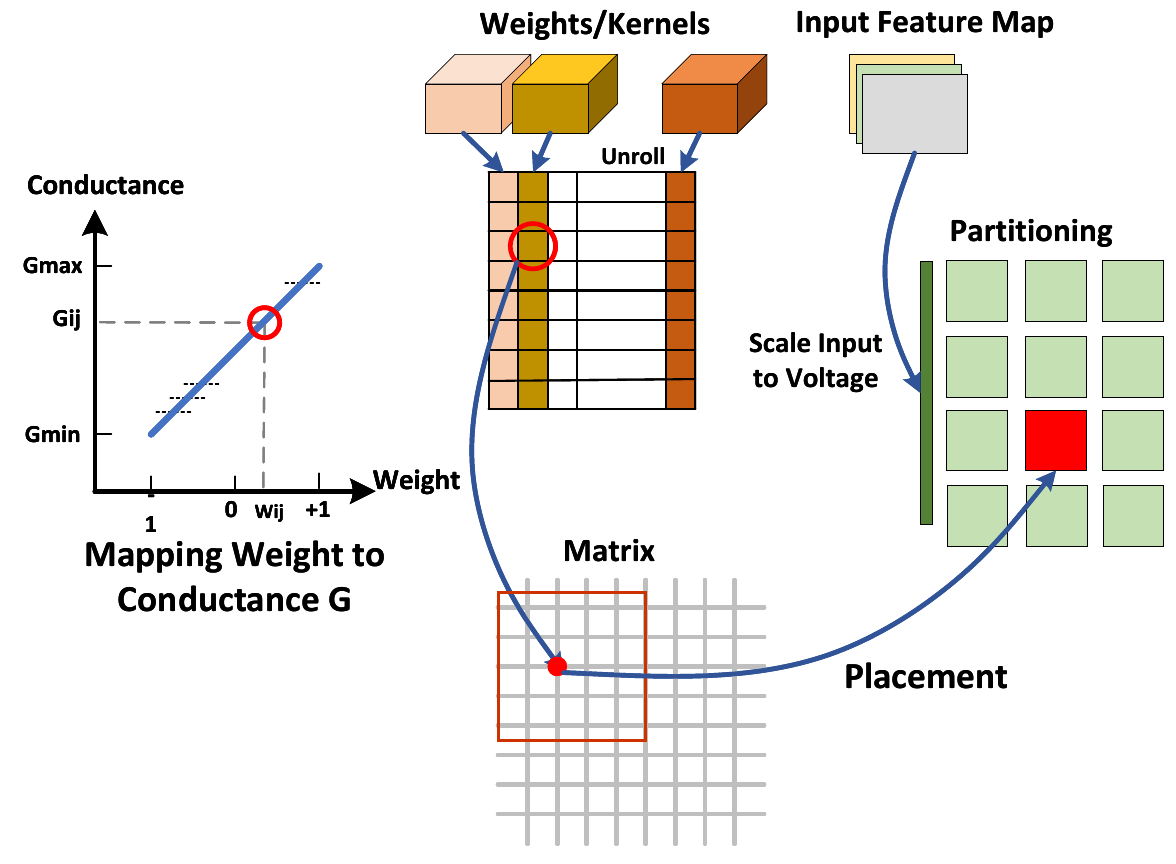}
	\caption{Overview of the mapping and placement process of learning parameters onto ReRAM-based Accelerator.}
	\label{Fig2-mapping}
\end{figure}

Fig.~\ref{Fig2-mapping} illustrates the overall procedure for mapping and placement of learning parameters, such as weight or kernel matrices and bias vectors, onto ReRAM-based PIM accelerators. Initially, the weights and biases from \textit{FC} layers, as well as the kernels and biases from \textit{CONV} layers, are transformed into suitable matrix formats to enable accurate VMM within the crossbar architecture. These formatted matrices are then mapped into corresponding conductance values for ReRAM cells and then placed in a suitable ReRAM crossbar, taking into account the hierarchy of crossbars, tiles, and memory banks. Learning parameters are mapped to conductance values of ReRAM crossbar cells as~\cite{Lammie2020}:
 \begin{equation}
	g_{i,j} = (g_{\text{max}} - g_{\text{min}}) \times \frac{w_{i,j} - w_{\text{min}}}{w_{\text{max}} - w_{\text{min}}} + g_{\text{min}}
	\label{Eq3}
\end{equation}
where, $w$ represents the weight/kernel parameters and $g$ denotes the conductance of the ReRAM cells. $g_{max}$ and $g_{min}$ are the maximum and the minimum conductance values, respectively.

\begin{figure}[!t]
	\centering
	\includegraphics[width=0.3\textwidth]{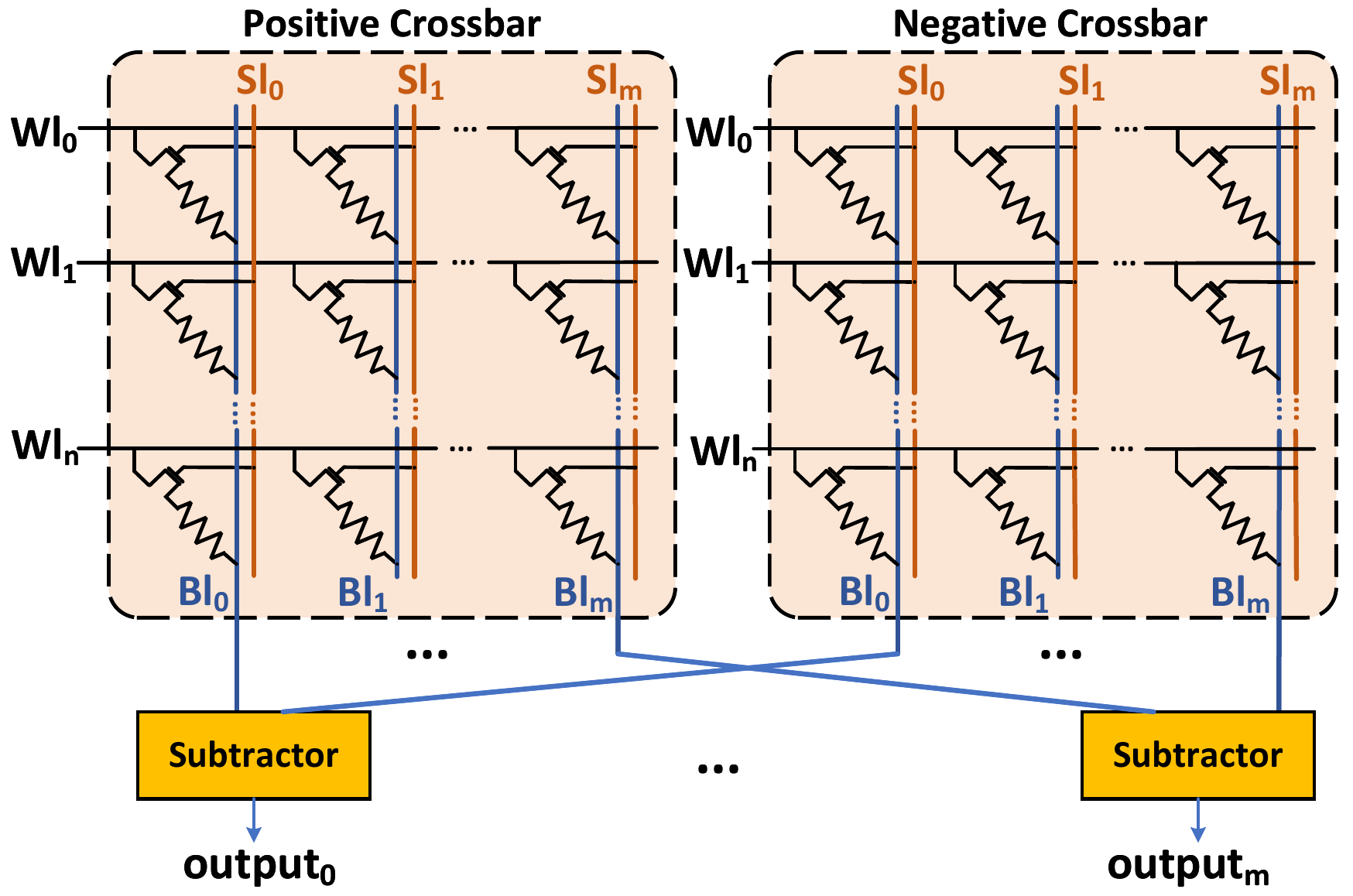}
	\caption{Dual crossbar architecture to store both positive and negative values of learning parameters.}
	\label{Fig3-dual-crossbar}
\end{figure}

Since the conductance of ReRAM cells is inherently non-negative, signed NN weights cannot be directly represented in a single crossbar. Therefore, a dual-crossbar structure consisting of separate positive and negative crossbars has been proposed~\cite{Chi2016, Sun2021} (as illustrated in Fig.~\ref{Fig3-dual-crossbar}). Each weight is encoded in differential form as $g_{i,j}=g_{i,j}^+-g_{i,j}^-$, where the absolute conductance value is programmed in the corresponding crossbar while the dual cell is set to $g_{min}$. The conductance of each cell is calculated by:
\begin{equation}
	\begin{aligned}
		g_{i,j}^+ &=
		\begin{cases}
			g_{i,j} & \text{if } w_{i,j} \geq 0 \\
			g_{\text{min}} & \text{if } w_{i,j} < 0
		\end{cases}
		\quad , \quad
		g_{i,j}^- =
		\begin{cases}
			g_{i,j} & \text{if } w_{i,j} \leq 0 \\
			g_{\text{min}} & \text{if } w_{i,j} > 0
		\end{cases}
	\end{aligned}
	\label{Eq4}
\end{equation}

\subsection{Related Work}
Several studies have investigated the reliability and vulnerability of CNNs to both permanent and transient faults across a variety of hardware platforms, including CPUs, GPUs, tensor processing units (TPUs), FPGAs, application-specific integrated circuits (ASICs)~\cite{Ibrahim2020, Gambardella2019, Zhang2019}, and ReRAM-based PIM accelerators~\cite{Dorostkar2024, Shin2023, Sawal2024}. To recover the degraded accuracy of CNNs caused by faults in ReRAM-based accelerators, several software- and hardware-based methods have been proposed.

Software-based methods typically focus on algorithmic strategies to mitigate the impact of both transient faults (e.g., resistance drift and process variation) and permanent faults. Notable methods include parameter mapping or remapping~\cite{ Liu2017, Shin2023, Ma2023, Huang2022, Tung2023}, retraining~\cite{Liu2015, Charan2020, Liu2017,Huang2022, Chen2017, Li2019RRAMedy}, fault- and variation-aware online training~\cite{Liu2015,Xia2019, Yeo2019, Wang2022, Dhingra2024}, and matrix transformation methods, e.g., row flipping, matrix permutation and value range transformation~\cite{ZhangBaogang2020}, all of which have been proposed to improve the reliability of the inference phase in NNs.  While these methods can improve fault tolerance without modifying the hardware architecture, they often introduce additional computational cost or memory overhead. In particular, retraining and online training approaches require frequent weight updates, resulting in a large number of write operations in ReRAM devices. Prior studies report that the number of training iterations for on-device training can reach on the order of $10^{4}-10^{8}$, while the average endurance of a memristor cell is about $10^{10}$ write cycles~\cite{ Song2021}, potentially accelerating device wear-out. Moreover, the effectiveness of software-based fault-tolerant methods significantly degrades under high fault rates and large device variations. Prior study reports that when the fault rate exceeds approximately 10\%-15\% of crossbar cells, the classification accuracy of CNN models may drop by more than 20\%–40\% with remapping methods~\cite{Bang2023}.

To specifically address faults in analog vector–matrix multiplication, several architecture-aware approaches have been proposed. For instance, a row shuffling method is introduced in~\cite{Zhang2020}, followed by post-processing computations to compensate the output errors. Several studies exploit the differential property of the hardware structure (the presence of two current paths used to generate signed weights, such as in dual-crossbar structure with positive and negative crossbars or differential column structures) as an opportunity for fault compensation without requiring additional hardware~\cite{Song2021, Shin2023, Bang2023}. In~\cite{Shin2023}, a fault-aware weight decomposition (FAWD) method is proposed to mitigate the impact of stuck-at faults (SAFs) in binary ReRAM crossbars, while~\cite{Bang2023} extends this approach to analog ReRAM crossbars. In addition, the forward parameter tuning (FPT) scheme proposed in~\cite{Quan2023} exploits the statistical properties of DNN computations to update network parameters in binary ReRAM devices. Although these approaches can reduce accuracy degradation caused by device-level faults, they often introduce additional computational overhead or require iterative parameter updates.

In contrast, hardware-based methods improve system reliability by introducing redundancy or modifying peripheral circuits ~\cite{Lee2017, Yeo2019, Xia2018}. A common approach is the use of error detection and correction codes (ECC)~\cite{Liu2018, Lou2020, Wang2015, Peng2021, Pinto2024, Liu2020}, which employ additional encoder and decoder circuits to detect and correct errors. These circuits introduce extra hardware complexity and may result in approximately 15\%-40\% additional area and power overhead, depending on the coding scheme and implementation, while improve accuracy about 50\%-60\%.  Redundancy can also be implemented through additional ReRAM cells~\cite{Xia2018}, redundant rows or columns, or redundant computation modules. For example, triple modular redundancy (TMR) replicates hardware modules three times and determines the final output using majority voting~\cite{ Lee2017}, which increases the hardware cost by nearly 200\% compared with the baseline design. Other redundancy-based methods employ redundant columns or crossbars to replace faulty ReRAM columns~\cite{Xia2018}. A distribution-aware redundancy scheme was proposed in~\cite{Xia2018}, which utilizes redundant ReRAM columns or crossbars to tolerate SAFs, while redundancy ratios ranging from approximately 200\% and 40\% to 60\% additional cells. However, applying redundancy in analog PIM accelerators is more challenging than in conventional memory designs. Consequently, conventional redundancy approaches, such as disabling entire rows or columns, may be too coarse-grained for analog computation and can lead to inefficiencies in area, energy consumption, and latency. Hardware compensation methods have also been proposed, such as modified transimpedance amplifier (TIA) circuits to compensate for SAF-induced error currents in ReRAM crossbar arrays~\cite{Huang2021, Xia2018}.

Despite extensive research on fault-tolerant techniques, most existing solutions are either designed for digital ReRAM crossbars or focus on memory-oriented applications. These approaches may not be directly suitable for analog ReRAM-based PIM accelerators that require accurate analog VMMs. Furthermore, purely algorithm-level methods often become less effective under high fault rates or large device variations, while hardware-based redundancy inevitably introduces additional hardware cost and power consumption. Therefore, combining hardware redundancy with software-based methods can provide a more balanced trade-off among reliability, hardware overhead, and classification accuracy.

In this work, we investigate the impact of redundancy ratio on classification accuracy, mean time to failure (MTTF), energy consumption, and area overhead in ReRAM-based PIM accelerators. We further optimize redundancy allocation to enhance the reliability of ReRAM crossbars in the presence of SAFs. In addition, a fault-aware mapping algorithm is proposed to further reduce the accuracy loss caused by SAFs. Compared with prior studies, the proposed scheme jointly considers redundancy design and fault-aware weight remapping, enabling more efficient and reliable computation in analog ReRAM-based accelerators. While prior work proposes valuable contributions, none provides a similar level of integration and efficiency in tolerating SAFs in analog ReRAM-based accelerators.

\section{ReSAFT: Fault-Tolerant Scheme for Analog ReRAM-Based PIM Accelerator}
\label{resaft}
\begin{figure}[!t]
	\centering
	\includegraphics[width=0.4\textwidth]{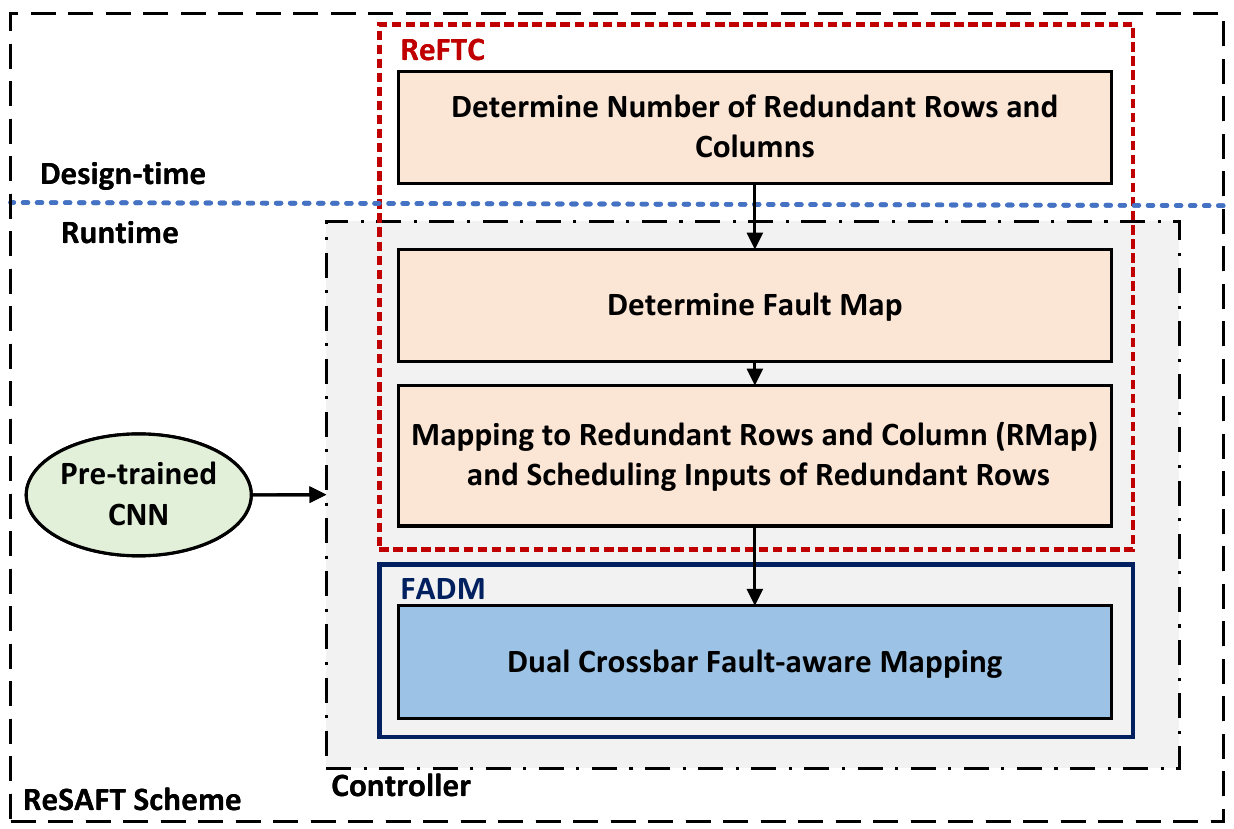}
	\caption{The overall flow of the ReSAFT scheme.}
	\label{Fig4-flowchart}
\end{figure}

The proposed \underline{S}tuck-\underline{A}t \underline{F}ault-\underline{T}olerant scheme for \underline{Re}RAM crossbar, called \textit{ReSAFT}, proposes a hardware- software-based solutions to tolerate SAFs occurs in ReRAM crossbars that performs analog VMMs for CNNs. ReSAFT utilize a redundancy-based architecture with runtime fault-aware mapping and remapping method to tolerate \textit{SaH} and \textit{SaL} faults efficiently.
The overall workflow of the proposed ReSAFT is illustrated in Fig.~\ref{Fig4-flowchart}. The \underline{re}dundancy-based \underline{f}ault-\underline{t}olerant ReRAM \underline{c}rossbar (\textit{ReFTC}) architecture integrates a combination of redundant rows and columns into the ReRAM crossbar structure to tolerate SAFs with minimal redundancy overhead. Unlike traditional methods that allocate redundancy heuristically, the ReFTC solution comprises two main stages: 1) employing a Multi-Objective Optimization (MOO) approach at design time to determine an adequate number of redundant rows and columns by considering reliability, MTTF, energy consumption, latency, area, and the classification accuracy of CNNs, and 2) Once the hardware configuration is fabricated and the specific fault map of each crossbar is identified, a mapping and scheduling algorithm (called RMap) is applied at runtime to efficiently utilize the available redundant rows and columns. The RMap maps faulty cells between redundant rows and columns and schedules the corresponding input voltages for redundant rows accordingly. 

To further enhance the fault tolerance of the accelerator, a \underline{F}ault-\underline{A}ware \underline{D}ual crossbar \underline{M}apping (\textit{FADM}) method is employed at runtime to remap the conductance value of faulty cells to their corresponding dual cells in the positive or negative ReRAM crossbars. we apply FADM specifically as a complementary and low-cost remapping method for the remaining faults that are not tolerated by the ReFTC architecture, as well as faults occurring in redundant rows and columns.

\subsection{ReFTC: Redundancy-Based Fault-Tolerant ReRAM Crossbar Architecture}
\begin{figure}[!t]
	\centering
	\includegraphics[width=0.35\textwidth]{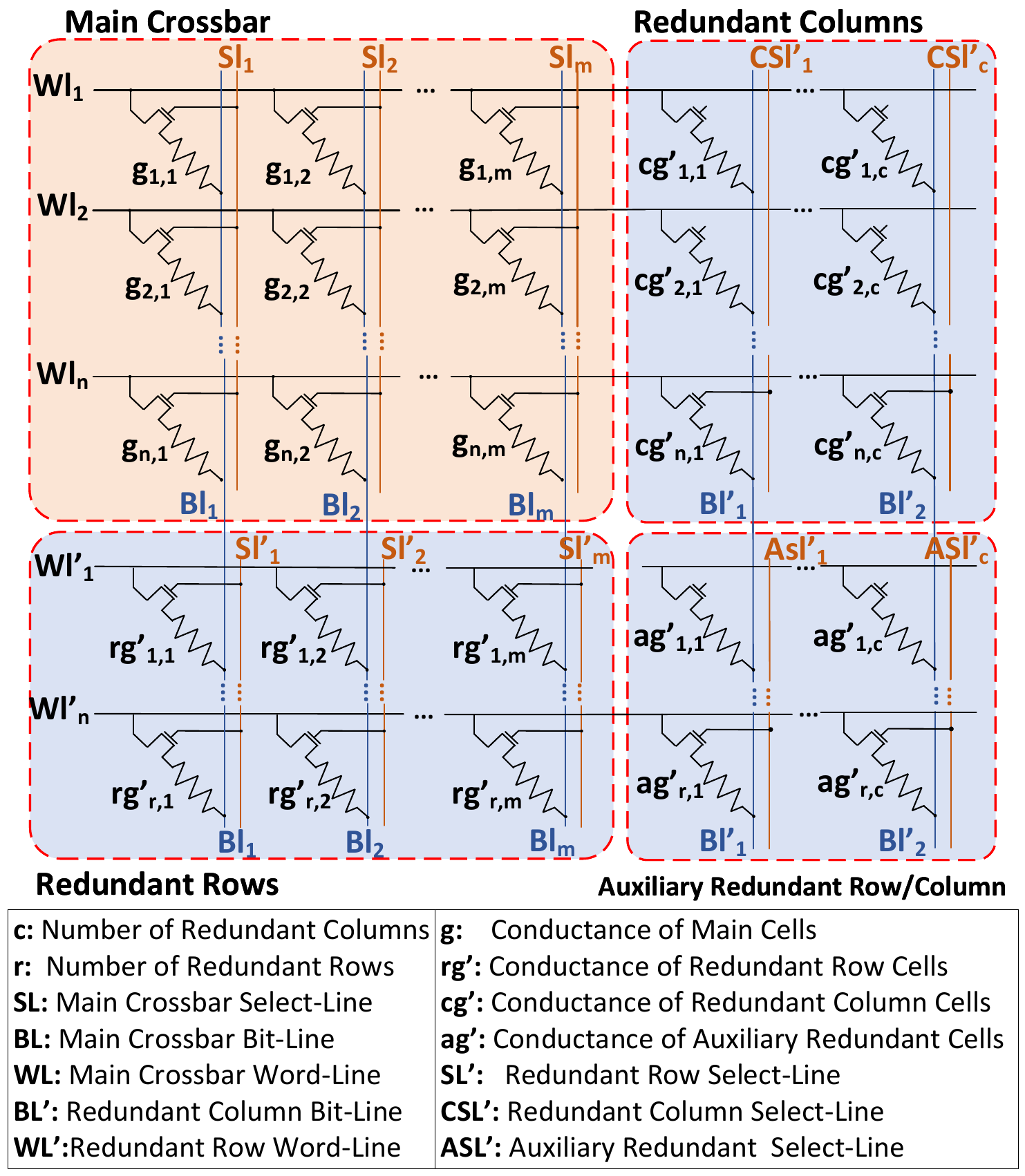}
	\caption{The proposed structure for redundancy-based fault-tolerant ReRAM crossbar (ReFTC).}
	\label{Fig5-crossbar-arrangment}
\end{figure}

The proposed redundancy-based fault-tolerant ReRAM crossbar architecture enhances the fault tolerance of ReRAM-based PIM accelerators against \textit{SaH} and \textit{SaL} faults by incorporating redundant rows and columns into the ReRAM crossbar architecture. The resulting design, referred to as the redundancy-based fault-tolerant ReRAM crossbar (\textit{ReFTC}), architecture is illustrated in Fig.~\ref{Fig5-crossbar-arrangment}. As depicted in the figure, the conventional \textit{1T1R} structure is employed as the basic cell for both the main crossbar and the redundant rows and columns. To tolerate faulty cells within the main crossbar, the architecture includes $r$ redundant rows and \textit{c} redundant columns. 
Furthermore, $r\times c$ auxiliary redundant cells are placed at the intersections of the redundant rows and redundant columns (as shown in Fig.~\ref{Fig5-crossbar-arrangment}) to tolerate possible faults occurring in the redundant columns themselves.

In the proposed ReFTC architecture, a combination of redundant rows and columns is employed to enhance fault tolerance. If only redundant columns are used in the architecture, a redundant column is required for each faulty column in the main crossbar. However, when multiple faults occur within a single column, all of them can be tolerated using just one redundant column. Conversely, if multiple faults occur in a single row of the main crossbar, multiple redundant columns would be needed to tolerate these faults. while, the same faults can be handled using just one redundant row. A similar principle applies when only redundant rows are utilized. If a high number of faults occur in a single column of the main crossbar, multiple redundant rows would be necessary, whereas a single redundant column would suffice. Therefore, our fault-tolerant architecture utilizing a combination of redundant rows and columns can achieve higher efficiency compared to using only redundant rows or columns.

Additionally, the redundant row, which can perform analog VMMs, a higher number of faults occurring in different rows of the main crossbar can be tolerated by introducing additional cycles with fewer additional rows. However, this method incurs a time overhead. For example, if five faults occur in various rows and columns of the main crossbar, all the faults can be tolerated for within five additional cycles using only one additional row. The more rows are added, the less significant this overhead becomes. However, if multiple faults occur in different columns of the main crossbar, there should be a sufficient number of additional columns corresponding to the number of faulty columns. If the number of additional columns is fewer than the number of faulty columns, it will not be possible to tolerate all faults, even with additional cycles.

\subsubsection{Mapping and Scheduling of Faulty Cells to Redundant Rows and Columns}
Prior to the mapping stage, a fault detection procedure is applied to the entire crossbar, including the redundant rows, redundant columns, and auxiliary cells, using algorithms such as March-C~\cite{Beckmann2016, Chen2015, VanDeGoor1994} and Squeeze-search~\cite{Beckmann2016} to generate a fault map.
If faults are detected in the crossbar, the proposed mapping method, called \textit{RMap}, is applied to map faulty cells of the main crossbar to the available redundant rows or columns, as well as to map faults in redundant columns to the available auxiliary redundant cells.	Using redundant columns, both \textit{SaH} and \textit{SaL} faults occurring in a column of the main crossbar can be tolerated. However, \textit{SaL} faults occurring in the main crossbar or redundant columns cannot be tolerated using redundant rows, since a large current flows through the faulty cell and cannot be mitigated by the redundant rows.
Therefore, redundant columns are prioritized for handling \textit{SaL} faults during mapping.

During the mapping stage, the proposed RMap algorithm considers not only faulty cells in the main crossbar but also faults occurring in the redundancy resources. If a fault is detected in a cell of a redundant column, the appropriate auxiliary cell is used as a replacement. Faults occurring in redundant rows can often be tolerated through the proposed mapping and scheduling algorithm across the remaining redundant rows. Auxiliary cells are primarily used when a faulty cell appears in the redundancy structure and cannot be handled through remapping. When auxiliary cells are not used for replacement, they are programmed to the minimum conductance value ($g_{min}$) to avoid contributing current during VMM operations.

In the proposed RMap method, the original conductance values, denoted as $g_{i,j}$, are mapped to redundant columns. The conductance values $g'_{i,j}$ of the redundant rows are configured such that the sum of the current from any column of the main crossbar ($I_{main}$) and the current from the corresponding redundant row $I_{redundant}$ match the golden current value, $I_{golden}$. The fault tolerated current of \textit{j-th} column ($I_{FT,j}$) is calculated by:
\begin{multline}
	\label{Eq5}
	I_{\text{ft},j} \approx I_{\text{main,j}} + I_{\text{redundant,j}} = 
	\sum_{i=0}^{n-1} V_i \cdot g_{i,j} + \sum_{i=0}^{n-1} V_i \cdot g'_{i,j}, \\
	\quad \forall j \in [0, m] \quad \text{where } g'_{i,j} \in \text{stuck-at cells}
\end{multline}

Therefore, if a \textit{SaH} fault occurs in a cell, the conductance value mapped to the corresponding redundant row cell is calculated by:
\begin{equation}
	\label{Eq6}
	rg'_{i,j} = \min \left\{ g_{i,j} - g_{\text{min}},\ g_{\text{min}} \right\}
\end{equation}
where $rg'_{i,j}$ represents the conductance of the redundant row for \textit{SaH} faults.
Moreover, a recent study of ReRAM crossbars reported that 9.04\% of the cells exhibit \textit{SaH} resistance states, whereas 1.75\% exhibit \textit{SaL} resistance states~\cite{Ching-YiChen2015}. This finding indicates that \textit{SaH} faults are more prevalent in ReRAM crossbar arrays compared to \textit{SaL} faults. 
Accordingly, the proposed ReFTC architecture prioritizes the use of redundant columns for tolerating columns containing \textit{SaL} faults, followed by columns with a high density of \textit{SaH} and \textit{SaL} faults, while redundant rows are primarily utilized for tolerating distributed \textit{SaH} faults.
If the number of available redundant rows and columns is insufficient to tolerate all detected faults, the remaining untolerated faults are further mitigated using the proposed FADM method. Since the redundant resources in ReFTC are limited, the proposed framework provides bounded fault tolerance depending on the allocated redundancy budget determined during the design-time optimization stage. Consequently, the effectiveness of fault tolerance depends on both the available redundancy resources and the remapping capability of FADM.

The proposed RMap method follows several steps to map faulty cells to redundant rows and columns:

\begin{enumerate}
\item{If the number of faulty cells within a specific column (row) of the main crossbar surpasses a predefined threshold, the conductance values of that entire column are mapped to a redundant column (row), and the original faulty column (row) is deactivated.}
\item{While there are unused redundant columns, columns with \textit{SaL} faults are deactivated and mapped to redundant columns.}
\item{In cases where the fault count does not exceed the threshold, \textit{SaH} faulty cells are instead mapped to redundant rows.}
\item{\textit{SaH} faulty cells are mapped to the same column in redundant rows as their original positions in the main crossbar, ensuring the index alignment is preserved.}
\item{Columns of redundant rows that contain no faulty cells are deactivated, and their cells are programmed to the minimum conductance value, $g_{min}$}.
\end{enumerate}

It is important to note that when the number of redundant rows is less than the number of rows in the main crossbar, faults from different rows of the main crossbar may be mapped to a single redundant row. In this scenario, different input voltage values must be applied to that row in successive cycles to ensure accurate computations, which causes serialization overhead. To mitigate the latency overhead of serialization, the RMap prioritizes redundant columns for mapping (Steps 1 and 2), and the algorithm maps only the remaining, distributed faults to redundant rows.
To efficiently manage the mapping of remaining faults in redundant rows, we propose a mapping and scheduling strategy that determines which input should be applied to each row during each cycle and which columns of the redundant rows should be activated. The proposed scheduling strategy is given in Algorithm~\ref{Algorithm1}.

This algorithm is designed to map and schedule \textit{SaH} faults from main crossbar rows to the available redundant rows. It begins by sorting the main crossbar rows in descending order based on the number of \textit{SaH} faults, obtaining both the sorted fault counts and their corresponding indices. Then, while there are still unmapped \textit{SaH} faults, the algorithm iteratively selects a main-crossbar row $x$, which has the highest number of unmapped faults (and is non-zero), and assigns it to \textit{cycle\_col} for the current redundant row \textit{r}. This value of $x$ serves as the input index for the current cycle (line 9). It then attempts to map faults from row \textit{x} to the redundant row $r$ (line 10 to 12). The input scheduling for each cycle is recorded in a list called \textit{cycle\_col}, which is later added to the overall cycle list to track the input sequence for each redundant row. If there are remaining faults and only a few rows left, a final mapping cycle is performed to ensure all faults are addressed (lines 14 to 17). The output includes: 1) \textit{redundant\_row}, a list of size $r\times m$ that indicates the mapping of \textit{SaH} faults in redundant rows and 2) cycle, a list that indicates the sequence of inputs that should be applied to each row in each cycle.

\begin{figure}[!t]
	\centering
	\includegraphics[width=0.35\textwidth]{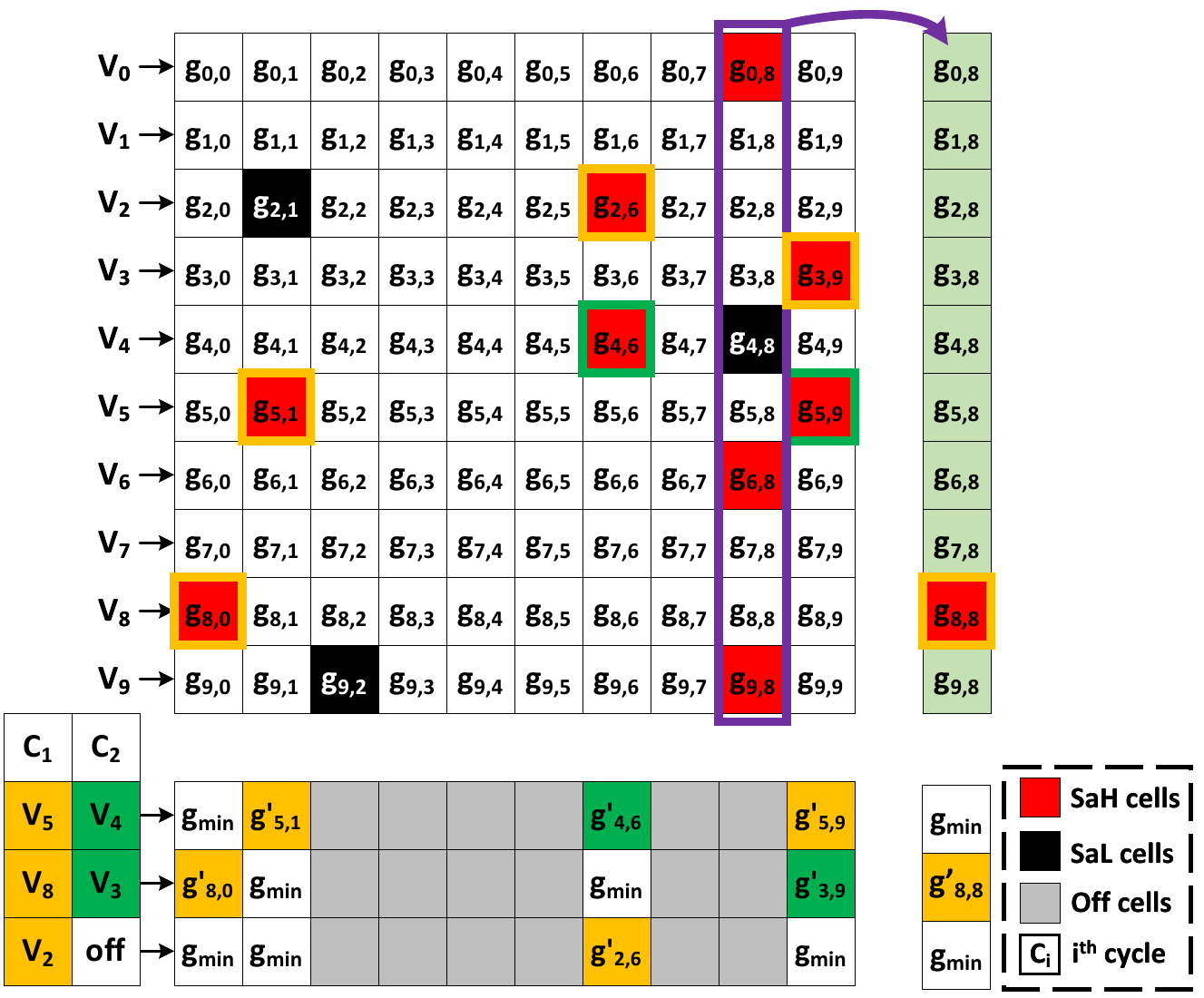}
	\caption{An example of the proposed redundancy-based fault tolerance structure.}
	\label{Fig6-example}
\end{figure}

\begin{algorithm}[t]
	\smaller
	\caption{Mapping and scheduling to redundant rows}
	\label{Algorithm1}
	\KwIn{Row, Number or redundant rows, fault map}
	\KwOut{redundant\_row (HRS fault mappings), cycle (to track the cycle mapping sequence)}
	\textit{redundant\_row} $\leftarrow$ a 2D list initialized with '\textit{off}' values\;
	\textit{cycle} $\leftarrow$ [ ]\;
	Sort number of \textit{SaH} faults in main rows in descending order, getting both sorted values and their corresponding indices\;
	\While{all \textit{SaH} faults are not yet mapped}{
		\textit{cycle\_col} $\leftarrow$ a list of '\textit{off}' values to represent the current mapping cycle\;
		\For{each r in row}{
			Retrieve the main row index \textit{x} with the next highest number of \textit{SaH} faults\;
			\If{the fault count is greater than zero}{
				\textit{cycle\_col}[r] $\leftarrow$ x\;
			}
		}
		\For{each y in column}{
			\If{an \textit{SaH} fault exists at position [x, y]}{
				redundant\_row[r, y] $\leftarrow$ [x, y]\;
			}
		}
		Append \textit{cycle\_col} to \textit{cycle}\;
		\If{there are still faults to map and only a few rows left}{
			Create a final mapping cycle \textit{cycle\_col\_last}\;
			Map and schedule the remaining faults similarly\;
			Append \textit{cycle\_col\_last} to \textit{cycle}\;
		}
	}
	\Return redundant\_row, cycle\;
\end{algorithm}
We explain the proposed structure with an example. Fig.~\ref{Fig6-example} shows a faulty $10\times10$ crossbar with three redundant rows and one redundant column. A total of ten \textit{SaH} faults and three \textit{SaL} faults are randomly distributed across the crossbar. The threshold for fault replacement is set to four. The column with index 8 in the main crossbar contains three \textit{SaH} faults and one \textit{SaL} fault.
All conductance values in this faulty column are remapped to the redundant column which contain one SaH, and the same input voltages corresponding to their original rows are applied. 
After mapping to the redundant column, five rows still contain \textit{SaH} and \textit{SaL} faults. Based on the mapping and scheduling procedure of Algorithm 1, these faults are mapped to cells in the redundant rows, while the fault in the redundant column is mapped to an auxiliary redundant cell.
Since the number of faulty rows is fewer than the number of redundant rows, faulty cells from multiple columns of the main crossbar can be mapped to a single redundant row. Therefore, different voltages are applied to the word-line of that redundant row in different cycles. In Fig.~\ref{Fig6-example}, the voltages $V_5$, $V_8$, and $V_2$ are applied to the word line of redundant rows in the first cycle, while the voltages $V_4$ and $V_3$ are applied in the second cycle.

The area and latency overheads of the proposed structure depend on the number of redundant rows and columns. In the worst case, the overhead can be in the order of $O(N)$, while in the best case, it can be reduced to $O(1)$.

\subsubsection{Determining Number of Redundant Rows and Columns}

The relationship between temporal and spatial redundancies as well as the number of rows and columns, involves a trade-off that should be carefully analyzed and optimized to balance the efficiency and accuracy. The goal is to select the number of redundant rows and columns that minimize energy consumption, latency, and area, while maximizing the MTTF. At the same time, the degradation in CNN accuracy due to faults should remain within a given threshold. This section presents a comprehensive analysis and model of how varying the number of redundant rows and columns affects reliability, MTTF, energy consumption, latency, area, and CNN classification accuracy. Based on this analysis, an optimization algorithm is proposed to determine the optimal redundancy configuration.

\paragraph{Reliability of ReRAM Crossbar}

Calculating the reliability of the proposed ReFTC is challenging because redundant rows and columns operate independently and simultaneously. If fault occurs, a redundant column replaces an entire faulty column. However, if the mapping involves a redundant row, only the faulty cell is mapped to one of the redundant cells in the same column as the location in the main crossbar, allowing both rows to remain active simultaneously. If redundant rows are fewer than main rows, multiple faults may share a single redundant row.

The ReFTC system can tolerate \textit{SaH} using either redundant rows or columns. However, \textit{SaL} faults can be tolerated only with redundant columns, which are exclusively effective for \textit{SaH} faults. The memory can operate as long as there are still usable cells in both main and redundant cells. Let \textit{n} and \textit{m} be the number of rows and columns in the ReRAM main crossbar, respectively, with \textit{c} redundant columns and \textit{r} redundant rows. The reliability of ReFTC depends on the reliability of redundant columns denoted as $R_{cols-FT}$ and redundant rows, denoted as $R_{rows-FT}$.

By considering redundant columns, the crossbar is reliable if at least m out of ($m+c$) columns, the total columns of ReFTC (the sum of main and redundant columns) are operational. Thus, the reliability of ReFTC based on the redundant columns can be expressed as:
\begin{equation}
	R_{\text{cols-FT}}(t) = \sum_{i = m}^{m+c} 
	\binom{m+c}{i} \bigg( R_{\text{col}}(t) \bigg)^i 
	\bigg( 1 - R_{\text{col}}(t) \bigg)^{m+c - i}
	\label{Eq7}
\end{equation}
where, $R_{\text{col}}$ is the reliability of a single column, which represents the probability that a column remains operational over time $t$. Each column contains $n$ independent cells with failure rate of $\lambda$. If a failure in any individual cell instantly causes the column to fail, the failure rate of a single column is $\lambda_{\text{col}} = n \lambda$ and the reliability of the column is $R_{\text{col}}(t) = e^{-\lambda_{\text{col}} t} = e^{-n \lambda t}$. If we assume that a column fails only when at least $j$ out of $n$ cells fail, the reliability of the column can be expressed using the binomial distribution. The column remains functional if fewer than $j$ cells fail. Therefore, the reliability of the column is given by:
\begin{equation}
	R_{\text{col-}j\ \text{cells fails}}(t) = 
	1 - \sum_{k = j}^{n} 
	\binom{n}{k} \left( 1 - e^{-\lambda t} \right)^k 
	\left( e^{-\lambda t} \right)^{n - k}
	\label{Eq8}
\end{equation}

MTTF is the expected value of time to fail, which is defined as $\text{MTTF}_{\text{col-FT}} = \int_0^\infty R_{\text{cols}}(t) \, dt$. MTTF of redundant columns when at least $m$ columns operate correctly can be calculated as~\cite{Kuo2003}:
\begin{equation}
	\text{MTTF}_{\text{col-FT}} = \frac{1}{\lambda_{\text{col}}} 
	\sum_{i = m}^{m + c} \frac{1}{i}
	\label{Eq9}
\end{equation}

By considering redundant rows, a maximum of $r$ faults in a column of the main crossbar can be tolerated, allowing the crossbar to function correctly. However, if the number of faults in any columns of the main crossbar exceeds $r$, these faults cannot be masked using the redundant columns, and the system will fail. Therefore, the reliability and MTTF of redundant rows can be calculated as: 
\begin{equation}
	R_{\text{rows-FT}}(t) = 
	\bigg( \sum_{k = 0}^{r} 
	\binom{n}{k} (1 - e^{-\lambda t})^k 
	( e^{-\lambda t})^{n - k} 
	\bigg)^m
	\label{Eq10}
\end{equation}
\begin{equation}
	\text{MTTF}_{\text{rows-FT}} = \int_{0}^{\infty} R_{\text{rows-FT}}(t)\, dt
	\label{Eq11}
\end{equation}

The MTTF of a single crossbar is given by:
$\text{MTTF}_{\text{crossbar}} = \text{MTTF}_{\text{rows}} + \text{MTTF}_{\text{cols}}$.
The MTTF of the whole system with $N$ crossbars is:
$\text{MTTF}_{\text{system}} = \min \left\{ \text{MTTF}_{\text{crossbar}, i} \right\}, \quad i = 1, 2, \dots, N$,
where $\text{MTTF}_{\text{crossbar}, i}$ denotes the MTTF of the $i-\text{th}$ crossbar.

When a CNN is deployed on ReRAM-based PIM accelerator, the system is considered to have failed if the CNN classification error in the presence of faults in the ReRAM crossbar cells exceeds a predefined threshold $\varepsilon$, compared to the error when the system is fault-free. Consequently, once all redundant rows or columns are fully utilized, it is no longer possible to tolerate additional faults, resulting in a degradation of the CNN’s accuracy. 

In order to validate the mathematical modeling of MTTF for this complex system, a Monte Carlo simulation was performed. Each simulation used the proposed ReFTC architecture with a $256\times256$ crossbar array and up to 256 redundant rows and columns. Both \textit{SaH} and \textit{SaL} faults were randomly injected into the crossbar using an exponential distribution with failure rates of $\lambda_{\text{SaH}} = \lambda_{\text{SaL}} = 10^{-5}$. The faulty cells were remapped to appropriate cells in the redundant rows and columns according to the proposed RMap method. The system was tracked over time to detect the earliest occurrence of an intolerable fault, i.e., when no redundant resources remained to tolerate new faults. MTTF was then calculated as the average failure time over 10,000 Monte Carlo iterations.

\begin{figure}[!t]
	\centering
	\subfloat[\label{Fig7-a}]{
		\includegraphics[width=0.45\linewidth]{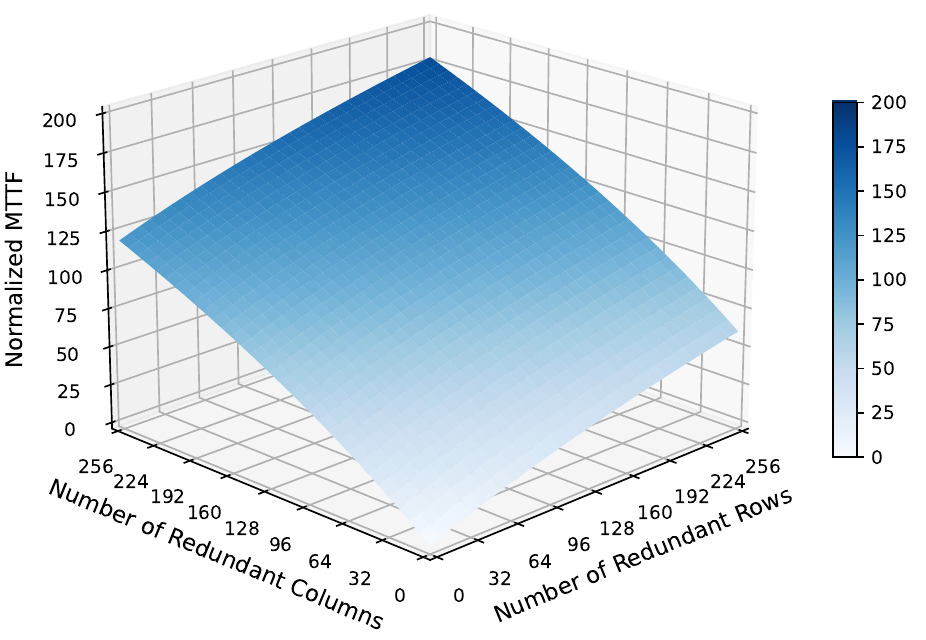}
	}
	\subfloat[\label{Fig7-b}]{
		\includegraphics[width=0.45\linewidth]{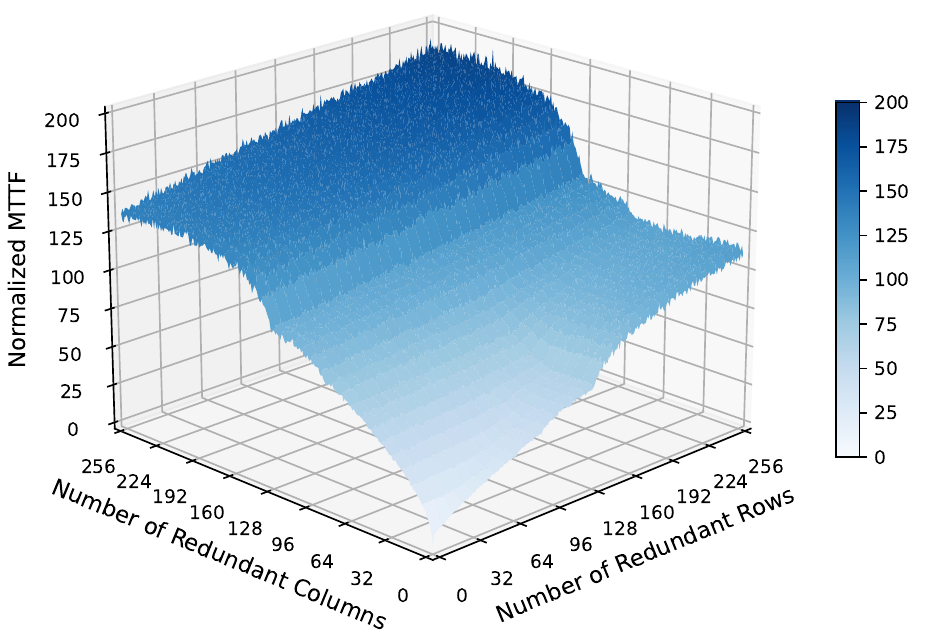}
	}
	\caption{MTTF of the ReFTC with different number of redundant rows/columns, normalized to that of a $256\times256$ main-crossbar, a) Mathemathical modeling, b) Monte Carlo Simulation.}
	\label{Fig7:multi2}
\end{figure}

Fig.~\ref{Fig7:multi2} shows the MTTF of the proposed ReFTC with different numbers of redundant rows and columns, normalized to a non-redundant $256 \times 256$ crossbar, evaluated using mathematical modeling and Monte Carlo simulation. Both mathematical modeling and Monte Carlo simulation demonstrate a consistent trend. As the number of redundant rows and columns increases, the MTTF improves significantly. The increase in MTTF is greater when adding redundant columns rather than redundant rows, because they can tolerate both \textit{SaH} and \textit{SaL} faults, while redundant rows can only tolerate \textit{SaH} faults.

The analytical surface in Fig.~\ref{Fig7-a} is smooth and continuous, reflecting the deterministic nature of the mathematical formulation. Compared to the analytical model, the Monte Carlo surface in Fig.~\ref{Fig7-a} is less smooth and exhibits local fluctuations due to the stochastic nature of the simulation. These variations are more significant in configurations with low redundancy, where the system is more sensitive to the fault distribution.

Monte Carlo simulation provides a more accurate estimation of the MTTF of the crossbar than the analytical model. A comparison between the analytical MTTF model and Monte Carlo simulation reveals a mean relative error (MRE) of 19.58\% and a Pearson correlation coefficient (PCC) of 0.9569. Although the high and close-to-1 PCC indicates a strong positive correlation between the two methods, the MRE is moderate and some visual deviations between the two curves are observable in Fig.~\ref{Fig7:multi2}, especially in low-redundancy configurations. These deviations reflect the approximation error in the analytical model. Nevertheless, the model successfully captures the overall trend and behavior of the system and offers sufficient accuracy for fast and scalable MTTF estimation.

\paragraph{Energy Consumption}

In general, adding redundant rows/columns increases the size of crossbar, which consequently leads to an increase in both its latency and energy consumption. The energy consumption of a crossbar can be calculated as the sum of the power consumption of read and write operations, multiplied by the latency of each operation, along with the leakage power during idle states as: 
\begin{equation}
	E_{\text{crossbar}} = 
	P_{\text{read}} \times \text{latency}_{\text{read}} + 
	P_{\text{write}} \times \text{latency}_{\text{write}} + 
	P_{\text{leakage}} \times T_{\text{idle}}
	\label{Eq12}
\end{equation}
where $P_{read}$, $P_{write}$, and $P_{leakage}$ represent the power consumption of read operations, write operations, and leakage power, respectively. Additionally, $latency_{read}$, $latency_{write}$, and $T_{idle}$ denote the latency of read and write operations and the idle state duration, respectively. The latency of read and write operations depends on the dimensions of the crossbar as well as the wiring delays. Therefore, as the crossbar size increases, the latency also increases due to the extended length of communication lines. In an accelerator operating in the inference phase, the number of write operations is typically much lower than that of read operations. Therefore, the majority of system power and energy is consumed by read operations. 

To compute the power consumption of read operations in a crossbar, the power of individual cells can be calculated and summed. The read operation power is generally derived by:
\begin{equation}
	P_{\text{read}} = \sum_{i=1}^{n_r} \sum_{j=1}^{m_r} V_i^2 \times G_{ij}
	\label{Eq13}
\end{equation}
where $V_i$ is the applied voltage to the row and $G_{ij}$ represents the conductance of the cell at row $i$ and column $j$. The variables $n_r$ and $m_r$ indicate the number of active rows and columns where read operations occur, respectively. To simplify the estimation of total power consumption in the crossbar during a read operation, an approximate model can be utilized as Eq.~\ref{Eq14}, assuming that all cells have an identical resistance ($R$) (equivalent resistance) and all rows receive an identical input voltage ($V$) (equivalent voltage). This behavioral model is not exact but can be employed to analyze and compare the impact of redundant rows and columns on energy consumption:
\begin{equation}
	P_{\text{read-est}} = n_r \times m_r \times \frac{V^2}{R}
	\label{Eq14}
\end{equation}

\begin{figure}[!t]
	\centering
	\subfloat[\label{Fig8-a}]{
		\includegraphics[width=0.45\linewidth]{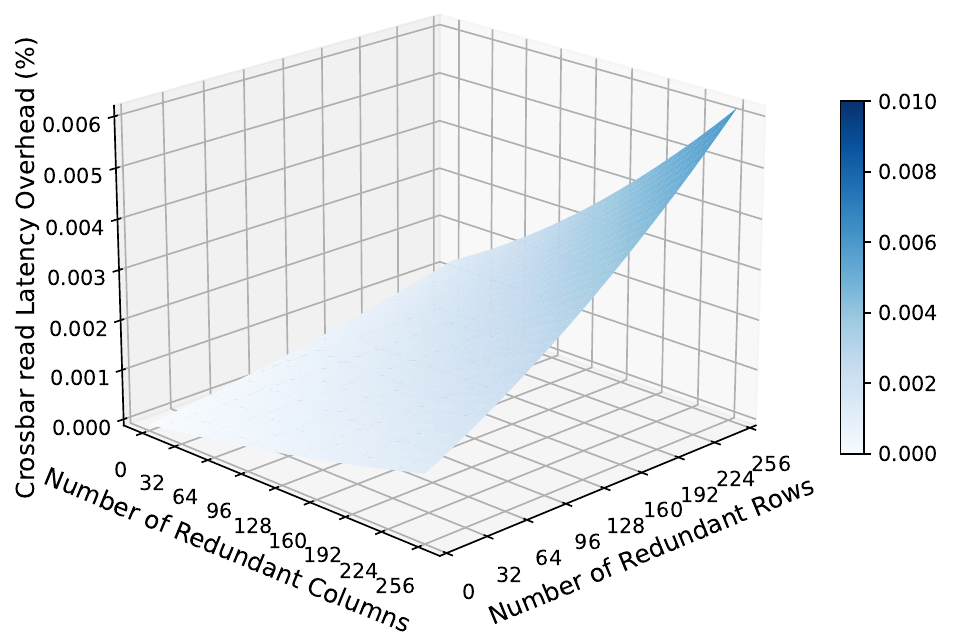}
	}
	\subfloat[\label{Fig8-b}]{
		\includegraphics[width=0.45\linewidth]{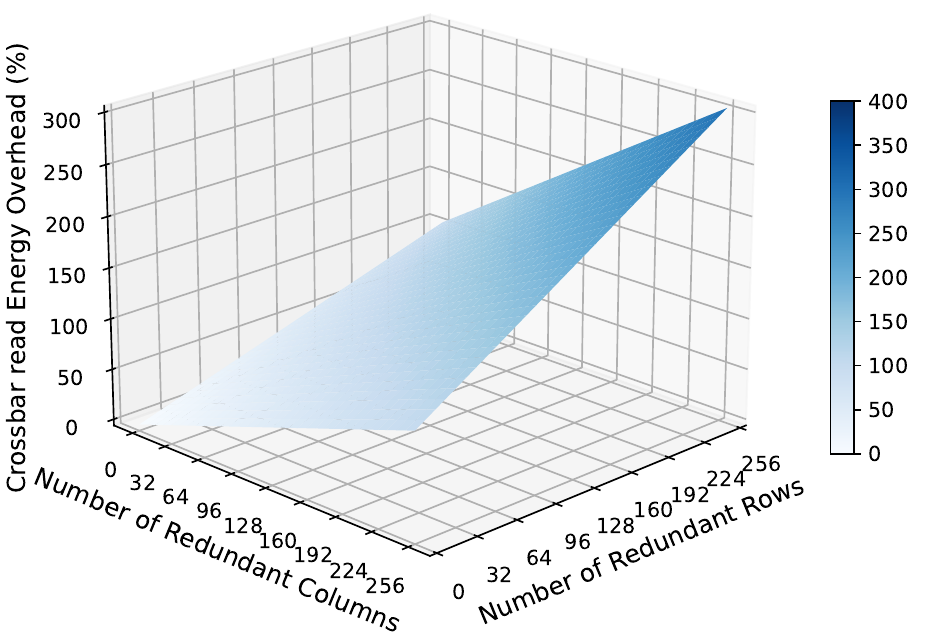}
	}
	\caption{The impact of the number of redundant rows and columns on a) the latency and b) the energy consumption of the read operation in a $256\times256$ main crossbar.}
	\label{Fig8:multi2}
\end{figure}

The power consumption equation for write operations is similar to that of read operations and can be generally derived as Eq.~\ref{Eq15} and be approximated as Eq.~\ref{Eq16}.
\begin{equation}
	P_{\text{write}} = \sum_{i=1}^{n_w} \sum_{j=1}^{m_w} V_{\text{prog},i}^2 \times G_{ij}
	\label{Eq15}
\end{equation}
\begin{equation}
	P_{\text{write-est}} = n_w \times m_w \times \frac{V_{\text{prog}}^2}{R}
	\label{Eq16}
\end{equation}
where $V_{prog}$ is the voltage applied to a cell during the write operation. The parameters $n_w$ and $m_w$ represent the number of cells in the rows and columns where write operations occur. 
The impact of the number of redundant rows and columns on the read latency and energy consumption of the read operation in a $256\times256$ main crossbar are shown in Figs.~\ref{Fig8-a} and~\ref{Fig8-b}, respectively.

\paragraph{Latency}

The number of redundant columns does not affect the number of cycles required for performing VMM operations. However, the number of redundant rows can alter the number of required cycles. Since cells from different rows of the main crossbar may be mapped to a single redundant row, inputs from two different rows of the main crossbar should be provided to that row in successive cycles. According to the fault distribution, as the number of redundant rows increases, the number of required cycles will decrease, while a smaller number of redundant rows may result in more required cycles. In the worst case, the maximum number of required cycles is roughly related to $\left\lceil \frac{n}{r} \right\rceil$, under the assumption of exclusive row redundancy.
However, in practice, the actual latency overhead is highly dependent on factors such as the number of redundant rows and columns, the fault distribution, and the efficiency of the mapping algorithm. By employing both row and column redundancy in the ReFTC architecture, and applying the RMap algorithm, faulty cells are distributed across both dimensions. Consequently, the need to map multiple faulty rows into a single redundant row is significantly reduced. Therefore, the real latency overhead is expected to be significantly lower than the worst-case bound.

\begin{figure}[!t]
	\centering
	\includegraphics[width=0.4\textwidth]{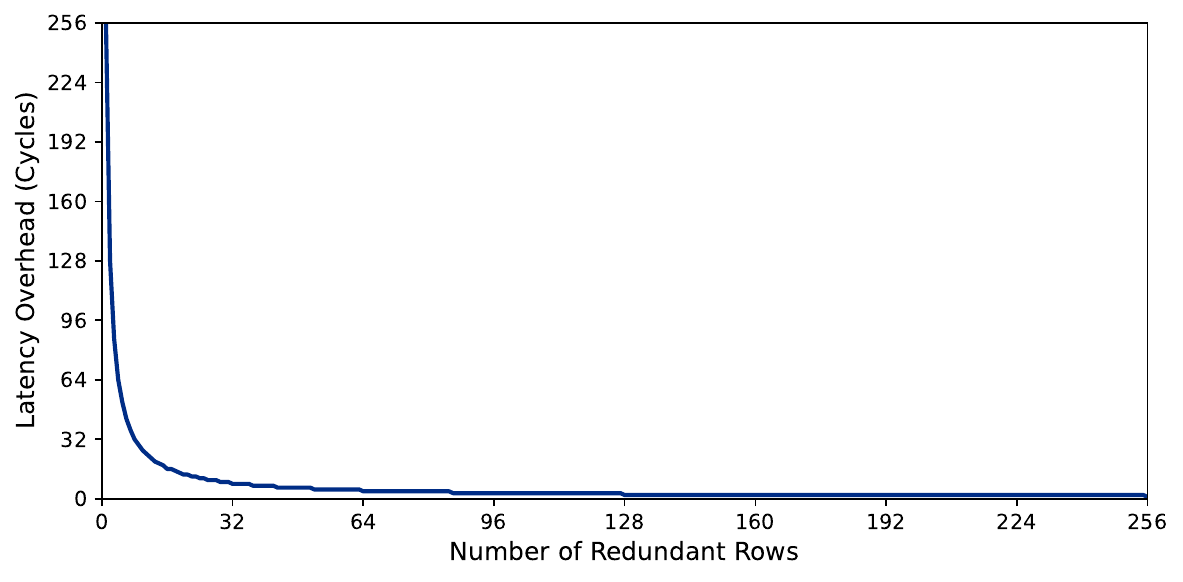}
	\caption{Worst-case latency overhead for different numbers of redundant rows in a $256\times256$ main crossbar.}
	\label{Fig9-Latency-overhead}
\end{figure}

The worst-case number of required cycles to perform VMM operation with different number of redundant rows in a $256\times256$ main crossbar is shown in Fig.~\ref{Fig9-Latency-overhead}. As shown in this figure, there is an inverse relationship between the number of redundant rows and the number of required cycles to perform VMM operations. A small number of redundant rows can drastically increase the latency overhead. In a $256\times256$ main crossbar with fewer than 16 redundant rows, more than 16 additional cycles are required to perform VMM operations. whereas, with more than 64 redundant rows, fewer than four additional cycles is added to performing VMM operation.

\paragraph{Area}

Let $n$ and $m$ be the number of rows and columns in the main crossbar, respectively. The total area of the ReRAM crossbar without redundancy can be calculated as:
\begin{equation}
	A_{\text{main-crossbar}} = n \times m \times A_{\text{cell}}
	\label{Eq17}
\end{equation}
where $A_{\text{cell}}$ is the area of a single memory cell, and for a \textit{1T1R} ReRAM structure, it can be estimated as~\cite{Zhu2023}:
\begin{equation}
	A_{\text{cell}} = 3\left(\frac{W}{L} + 1\right) \times F^2
	\label{Eq18}
\end{equation}
where $W/L$ is the transistor technology parameter, and $F$ is the feature size. By adding $r$ redundant rows and $t$ redundant columns, the total area of the FT-crossbar can be calculated as Eq.~\ref{Eq19}. The redundant rows and columns occupy the same area per cell as the main crossbar cells.

\begin{equation}
	A_{\text{redundant}} = \left[ (r \times n) + (c \times m) + (r \times c) \right] \times A_{\text{cell}}
	\label{Eq19}
\end{equation}

\begin{figure}[!t]
	\centering
	\includegraphics[width=0.3\textwidth]{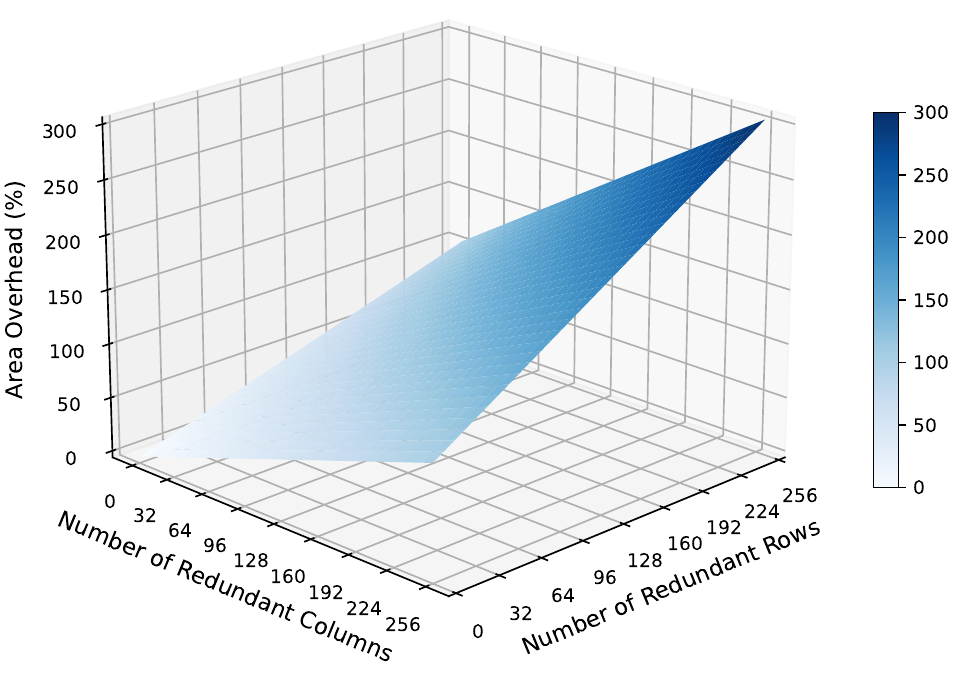}
	\caption{Area overhead for different numbers of redundant rows and columns in a $256\times256$ main crossbar.}
	\label{Fig10-area-overhead}
\end{figure}

The total area of the crossbar memory, including both the main crossbar and the redundancy, is the sum of the areas calculated by Eqs.~\ref{Eq17} and~\ref{Eq19}. In addition to the memory cells themselves, there is also some area overhead associated with logic circuits that manage the remapping of faulty rows or columns to redundant ones. The area overhead with different number of redundant rows and columns in a $256\times256$ main crossbar is shown in Fig.~\ref{Fig10-area-overhead}. 

\paragraph{Accuracy}
Top-1 accuracy measures the proportion of correct predictions and is a proper metric to evaluate the impact of faults on CNNs and the effectiveness of the ReSAFT scheme under faulty conditions.

Exhaustively evaluating the accuracy of CNNs under all possible combinations of redundant rows and columns is time and resource expensive. For example, for a $256\times256$ main crossbar, considering 257 possible values (ranging from 0 to 256) for each redundancy types leads to over $65,000$ configurations. Computing the accuracy of CNNs for all these combinations is not practical, especially for deeper networks or larger crossbars. To address this challenge, we first evaluated the classification accuracy of CNNs under varying numbers of redundant rows and columns separately, and fitted two independent models, $Accuracy_{row} (r)$ and $Accuracy_{col} (c)$, respectively. This approach required only 514 simulations in total (257 for rows and 257 for columns), in contrast to over $65,000$ simulations needed for all possible combinations. However, these one-dimensional models cannot directly estimate the accuracy for arbitrary $(r,c)$ combinations. Therefore, we propose an approximation model in which the total redundancy $r+c$ is used as the input to both functions, and their outputs are averaged as:
\begin{equation}
	\text{Accuracy}_{\text{FT}}(r, c) = 
	\frac{\text{Accuracy}_{\text{row}}(r + c) + \text{Accuracy}_{\text{col}}(r + c)}{2}
	\label{Eq20}
\end{equation}

This approach provides a good balance between accuracy and computational efficiency, reducing the need for exhaustive simulations while maintaining acceptable estimation error.

\subsubsection{Optimization Problem}
In this section, an optimization algorithm is presented to determine the optimal values for the number of redundant rows ($r$) and columns ($c$), aiming to optimize multiple objective functions, including MTTF, energy consumption, area, and latency. This problem falls into the category of MOO, as it involves conflicting objectives: while energy consumption, latency, and area should be minimized, the MTTF, which corresponds to the minimum MTTF among all crossbars, should be maximized.

Additionally, classification accuracy is considered as a constraint, ensuring that the accuracy deviation between the proposed redundancy-based fault-tolerant accelerator ($Accuracy_{FT}$) and the baseline classification accuracy ($Accuracy_{Baseline}$) remains less than a predefined threshold $\varepsilon$. The value of $\varepsilon$ depends on the CNN model and its application, directly affecting the number of optimal solutions. The range of redundant rows and columns is defined as $[0,n]$ and $[0,m]$, respectively. We can formulate this problem as Eqs.~\ref{Eq21} to~\ref{Eq25}.
\begin{equation}
	\min \text{Energy}_{\text{system}} = \sum_{i=1}^{N_{\text{crossbar}}} \text{Energy}_{\text{crossbar},i}
	\label{Eq21}
\end{equation}
\begin{equation}
	\min \text{Latency}_{\text{system}} = \max \left\{ \text{Latency}_{\text{crossbar},i} \right\}, \quad 1 < i < N_{\text{crossbar}}
	\label{Eq22}
\end{equation}
\begin{equation}
	\min \text{Area}_{\text{system}} = N_{\text{crossbar}} \times \text{Area}_{\text{crossbar}}
	\label{Eq23}
\end{equation}
\begin{equation}
	\max \text{MTTF}_{\text{system}} = \min \left\{ \text{MTTF}_{\text{crossbar},i} \right\}, \quad 1 < i < N_{\text{crossbar}}
	\label{Eq24}
\end{equation}
\begin{equation}
	\begin{aligned}
		&\left| \text{Accuracy}_{\text{Baseline}} - \text{Accuracy}_{\text{FT}} \right| \leq \varepsilon \\
		&0 \leq r \leq n,\quad r \in \mathbb{N} \\
		&0 \leq c \leq m,\quad c \in \mathbb{N}
	\end{aligned}
	\label{Eq25}
\end{equation}

\subsection{FADM: Fault-Aware Weight Mapping in Dual ReRAM Crossbar}

In the dual-crossbar architecture, which is originally employed to store positive and negative weight values, faulty ReRAM cells can also be tolerated using their corresponding cells in the opposite crossbar. Since two types of faults (\textit{SaH} and \textit{SaL}) may occur in both positive and negative crossbars, each crossbar can be in one of three states: fault-free, \textit{SaH}, or \textit{SaL}. Therefore, a pair of dual crossbars can experience nine possible fault cases. 
Among these cases, one corresponds to the fault-free condition, four cases occur when faults affect a cell in both crossbars simultaneously (representing double faults) and are therefore intolerable. The remaining four cases involve a fault in only one crossbar (representing single faults) and can be tolerated using the proposed FADM method by exploiting the complementary dual-crossbar structure.
The proposed FADM method is particularly effective for tolerating \textit{SaL} faults in the positive crossbar, \textit{SaH} faults in the negative crossbar, as well as faults located in redundant rows and columns.

\begin{figure*}[!t]
	\centering
	\subfloat[\label{Fig11-a}]{
		\includegraphics[width=0.1\linewidth]{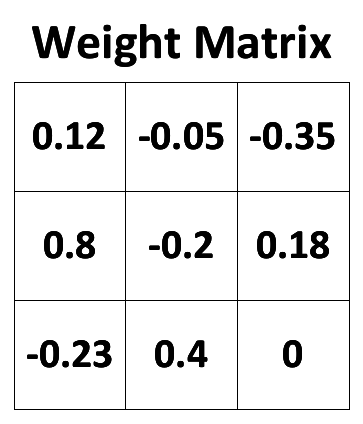}
	}
	\subfloat[\label{Fig11-b}]{
		\includegraphics[width=0.35\linewidth]{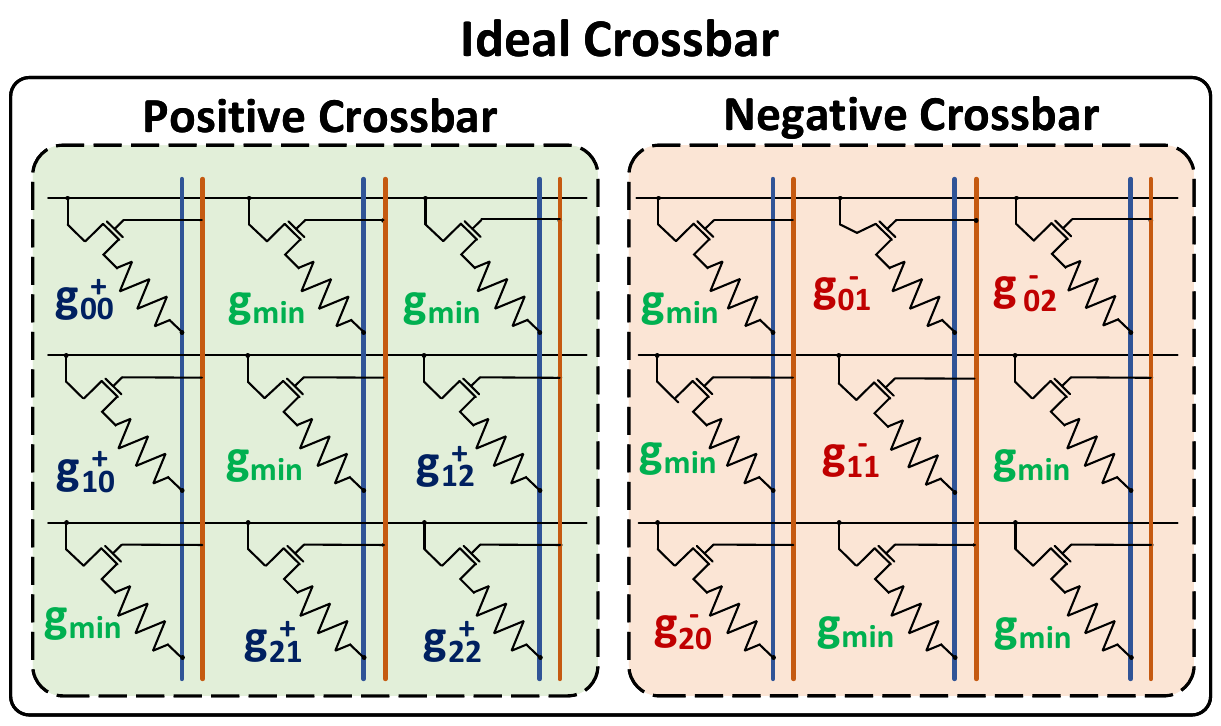}
	}
	\subfloat[\label{Fig11-c}]{
		\includegraphics[width=0.354\linewidth]{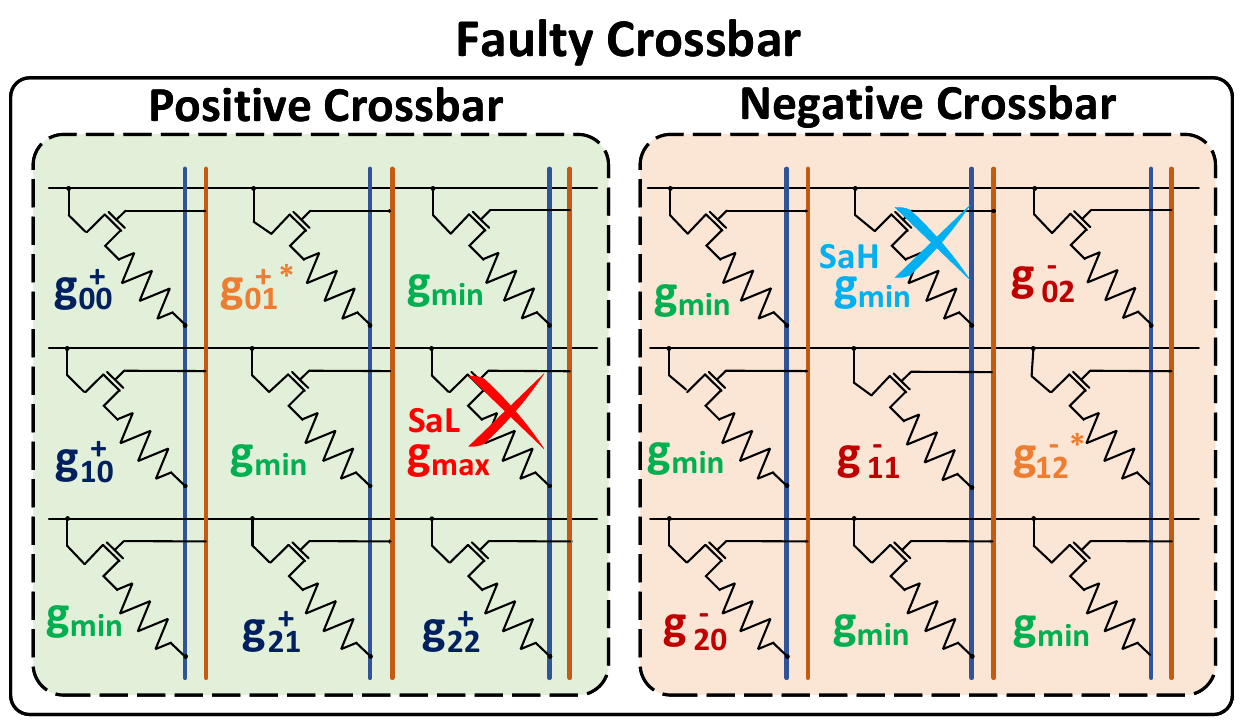}
	}
	\caption{An example of fault-aware weight mapping (FADM) in a dual crossbar, (a) weight matrix, (b) ideally mapped dual-crossbar, and (c) fault-tolerant dual-crossbar for \textit{SaL} and \textit{SaH} in positive and negative crossbars. respectively.}
	\label{Fig11:multi3}
\end{figure*}

For the weight matrix illustrated in Fig.~\ref{Fig11-a}, the ideal (fault-free) condutance mapping in dual crossbars are shown in Fig.~\ref{Fig11-b}. Meanwhile, Fig.~\ref{Fig11-c} demonstrates the tolerance of \textit{SaL} faults in the positive crossbar and \textit{SaH} in negative crossbar, achieved by utilizing their corresponding cells in the opposite crossbar.

Eqs.~\ref{Eq26} to~\ref{Eq29} calculate the required conductivity values for the proposed mapping to mitigate the effect of faults. In these equations, $g_{(i,j)}^{+}$ and $g_{(i,j)}^{-}$ represent the original conductivity values that should be programmed in cell $(i,j)$ of the positive and negative crossbars, respectively, in the absence of faults. Conversely, ${g_{(i,j)}^{+}}^{*}$ and ${g_{(i,j)}^{-}}^{*}$ denote the fault-aware conductivity values that should be programmed in the corresponding cells when faults occur. It should be noted that these conductance values should fall within the range of $[g_{\text{min}}, g_{\text{max}}]$.

In the first case, if cell $(i,j)$ in positive crossbar is in the \textit{SaH} resistance state, then the conductance value in that cell will be equal to minimum conductance value ($g_{(i,j)}^{+*} = g_{\text{min}}$). The fault-aware value that should be mapped to relevant cell in negative crossbar will be given as:
\begin{equation}
	g_{(i,j)}^{-*} = g_{(i,j)}^{-} + g_{\text{min}} - g_{(i,j)}^{+}
	\label{Eq26}
\end{equation}

In the second case, if cell $(i,j)$ in the positive crossbar is in the \textit{SaL} resistance state, then the conductance value in that cell will be equal to maximum conductance value ($g_{(i,j)}^{+*} = g_{\text{max}}$). The fault-aware value that should be mapped to relevant cell in negative crossbar will be given as:
\begin{equation}
	g_{(i,j)}^{-*} = g_{(i,j)}^{-} + g_{\text{max}} - g_{(i,j)}^{+}
	\label{Eq27}
\end{equation}

In the third case, if cell $(i,j)$ in negative crossbar is in the \textit{SaH} resistance state, then the conductance value in that cell will be equal to minimum conductance value ($g_{(i,j)}^{-*} = g_{\text{min}}$). The fault-aware value that should be mapped to relevant cell in positive crossbar will be given as:
\begin{equation}
	g_{(i,j)}^{+*} = g_{(i,j)}^{+} + g_{\text{min}} - g_{(i,j)}^{-}
	\label{Eq28}
\end{equation}

In the fourth case, if cell $(i,j)$ in negative crossbar is in the \textit{SaL} resistance state, then the conductance value in that cell will be equals to maximum conductance value ($g_{(i,j)}^{-*} = g_{\text{max}}$). The fault-aware value that should be mapped to relevant cell in positive crossbar will be given as:
\begin{equation}
	g_{(i,j)}^{+*} = g_{(i,j)}^{+} + g_{\text{max}} - g_{(i,j)}^{-}
	\label{Eq29}
\end{equation}

Depending on whether the NN weight is positive or negative, and the location of \textit{SaH} and \textit{SaL} faults in the positive or negative crossbar, the faults can either be tolerated or not. The corresponding fault-aware conductance values for the positive and negative crossbar cells are presented in Table~\ref{Table1:fault-aware-mapping}. When the weight is positive and an \textit{SaH} occurs in the positive crossbar cell, the value of Eq.~\ref{Eq26} will be less than $g_{\text{min}}$, and it cannot be tolerated. Similarly, in the intolerable states, when the weight is positive and a \textit{SaL} fault occurs in the negative crossbar cell, or when the weight is negative and an \textit{SaL} fault occurs in the positive crossbar cell, or when an \textit{SaH} fault occurs in the negative crossbar cell, or in cases where the corresponding cells in both crossbars experience faults, these faults cannot be tolerated.
Symmetric intolerable double fault cases, such as (\textit{SaH}, \textit{SaH}) and (\textit{SaL}, \textit{SaL}), functionally act as ideal states when the target weight is zero ($w_{i,j}=0$).

\begin{table*}[t]
	\centering
	\caption{Fault-aware conductance values in the positive and negative crossbar cells}
	\begin{adjustbox}{max width=1\linewidth}
		\begin{tabular}
			{
				>{\centering\arraybackslash}m{2cm}
				>{\centering\arraybackslash}m{4cm}
				>{\centering\arraybackslash}m{5cm}
				>{\centering\arraybackslash}m{5cm}
				>{\centering\arraybackslash}m{3cm}
			}
			\hline
			Weight value & Fault location & Positive crossbar conductance & Negative crossbar conductance & State \\
			\hline
			\multirow{5}{*}{$W>0$} 
			& Fault-free                         & \cellcolor{cyan!15}$g_{(i,j)}^+$ & \cellcolor{cyan!15}$g_\mathrm{min}$ & \cellcolor{cyan!15}Ideal \\
			& \textit{SaH} in the positive crossbar      & \cellcolor{red!15}$g_\mathrm{min}$ & \cellcolor{red!15}$g_{(i,j)}^{-*} = g_\mathrm{min}$ & \cellcolor{red!15}Intolerable faults \\
			& \textit{SaL} in the positive crossbar      & $g_\mathrm{max}$ & $g_{(i,j)}^{-*} = g_\mathrm{max} - g_{(i,j)}^+ + g_\mathrm{min}$ & Tolerable faults \\
			& \textit{SaH} in the negative crossbar      & $g_{(i,j)}^{+*} = g_{(i,j)}^+$ & $g_\mathrm{min}$ & Tolerable faults \\
			& \textit{SaL} in the negative crossbar      & \cellcolor{red!15}$g_{(i,j)}^{+*} = g_\mathrm{max}$ & \cellcolor{red!15}$g_\mathrm{max}$ & \cellcolor{red!15}Intolerable faults \\
			\hline
			\multirow{5}{*}{$W<0$} 
			& Fault-free                         & \cellcolor{cyan!15}$g_\mathrm{min}$ & \cellcolor{cyan!15}$g_{(i,j)}^-$ & \cellcolor{cyan!15}Ideal \\
			& \textit{SaH} in the positive crossbar      & $g_\mathrm{min}$ & $g_{(i,j)}^{-*} = g_{(i,j)}^-$ & Tolerable faults \\
			& \textit{SaL} in the positive crossbar      & \cellcolor{red!15}$g_\mathrm{max}$ & \cellcolor{red!15}$g_{(i,j)}^{-*} = g_\mathrm{max}$ & \cellcolor{red!15}Intolerable faults \\
			& \textit{SaH} in the negative crossbar      & \cellcolor{red!15}$g_{(i,j)}^{+*} = g_\mathrm{min}$ & \cellcolor{red!15}$g_\mathrm{min}$ & \cellcolor{red!15}Intolerable faults \\
			& \textit{SaL} in the negative crossbar      & $g_{(i,j)}^{+*} = g_\mathrm{max} + g_\mathrm{min} - g_{(i,j)}^-$ & $g_\mathrm{max}$ & Tolerable faults \\
			\hline
		\end{tabular}
	\end{adjustbox}
	\label{Table1:fault-aware-mapping}
\end{table*}

Although FADM does not provide high fault coverage, it significantly improves reliability with negligible hardware overhead. The reason for proposing FADM is to combine it with the ReFTC architecture (and other methods) to improve accuracy while reducing both hardware and software overhead.

\section {Evaluation and Analysis}
\label{evaluation}
\subsection {Experiment Setup}
To evaluate the performance of our proposed methods, we conducted a set of experiments using well-known CNN models, including \textit{SimpleNet}~\cite{SimpleNet}, \textit{LeNet5}~\cite{Lecun1998}, \textit{VGG8}~\cite{Simonyan2015}, \textit{VGG11}~\cite{Simonyan2015}, and \textit{ResNet18}~\cite{He2016}. \textit{SimpleNet} and \textit{LeNet5} were trained on the \textit{MNIST} dataset, which consists of handwritten digits, while \textit{VGG8} and \textit{ResNet18} were trained on the \textit{CIFAR-10}~\cite{krizhevsky2009learning} dataset, a collection of small, color images. Both \textit{MNIST}~\cite{Lecun1998} and \textit{CIFAR-10} datasets contain 60,000 images for training and 10,000 images for testing, classified into ten different classes. The specifications of the CNN models evaluated in this study are presented in Table~\ref{Table2:cnn-characteristics}. \textit{PyTorch} with the \textit{MemTorch} library are used to model ReRAM-based DNNs and evaluate classification accuracy of faulty and non-faulty crossbars. All ReRAM devices were simulated using the \textit{VTEAM} model~\cite{Kvatinsky2015}, with $R_{ON} = 1.4\times10^4$ and $R_{OFF} = 1.4\times10^7$.

To estimate metrics such as accuracy, energy consumption, latency, and area utilization, the ReRAM-based PIM architecture was built using \textit{MNSIM2.0}~\cite{Zhu2023}, an open-source behavioral-level modeling tool for memristor-based PIM architectures. We modified the crossbar architectures and add redundant rows/columns and peripheral circuit. For the ReRAM-based crossbar array (RCA) configuration, we adopted a $256\times256$ dual-crossbar tile structure. This configuration includes positive and negative crossbars to store the positive and negative CNN learning parameters, respectively. The RCA was implemented using a \textit{1T1R} structure with redundant rows and columns, as well as CMOS-based peripheral circuits, fabricated using TSMC 45nm device technology.

Moreover, to solve the proposed MOO problem, we use the Non-dominated Sorting Genetic Algorithm-II (NSGA-II)~\cite{Deb2002} algorithm, and its implementation was conducted using the Python Multi-Objective Optimization (PYMOO) framework~\cite{Blank2020}, a dedicated \textit{Python} library for multi-objective optimization. PYMOO provides a comprehensive set of tools for analyzing and managing trade-offs among competing objectives.

\begin{table*}[!t]
	\centering
	\caption{Specifications of the CNN models under study}
	\begin{adjustbox}{max width=1\linewidth}
		\begin{tabular}{
				>{\centering\arraybackslash}m{2cm}
				>{\centering\arraybackslash}m{2cm}
				>{\centering\arraybackslash}m{3cm}
				>{\centering\arraybackslash}m{2.5cm}
				>{\centering\arraybackslash}m{3cm}
				>{\centering\arraybackslash}m{2.5cm}
				>{\centering\arraybackslash}m{2.5cm}
				>{\centering\arraybackslash}m{2.5cm}
			}
			\hline
			CNNs & Dataset & Baseline Accuracy & Accuracy in HW & \# of CONV Layers & \# of FC Layers & Total Parameters & Total Size (\textit{MB}) \\
			\hline
			SimpleNet & MNIST & 99.42\% & 99.36\% & 2 & 2 & 431{,}080 & 1.76 \\
			LeNet5    & MNIST & 99.26\% & 99.19\% & 2 & 3 & 44{,}426  & 0.3 \\
			VGG8      & Cifar-10 & 93.45\% & 93.36\% & 6 & 3 & 13{,}499{,}530 & 62.54 \\
			VGG11     & Cifar-10 & 86.14\% & 85.39\% & 8 & 1 & 9{,}231{,}114 & 38.94 \\
			ResNet18 & Cifar-10 & 89.42\% & 88.76\% & 20 & 1 & 11{,}173{,}962 & 53.89 \\
			\hline
		\end{tabular}
	\end{adjustbox}
	\label{Table2:cnn-characteristics}
\end{table*}

\subsection{Classification Accuracy of CNN Models on ReFTC Architecture}
\begin{figure*}[!t]
	\centering
	\subfloat[\label{Fig12-a}]{
		\includegraphics[width=0.45\linewidth]{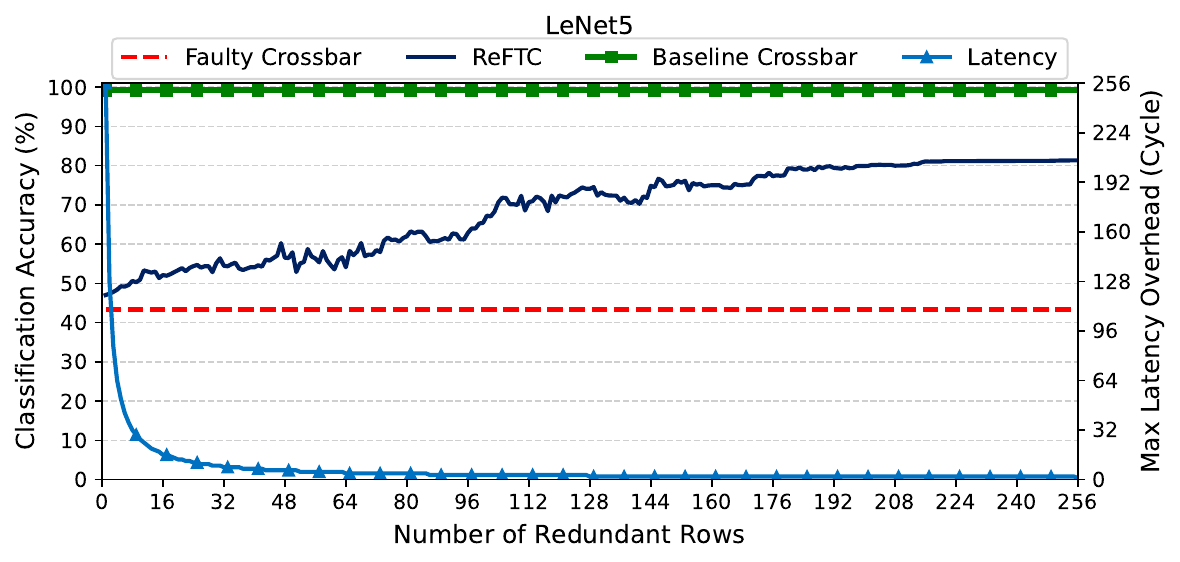}
	}
	\subfloat[\label{Fig12-b}]{
		\includegraphics[width=0.45\linewidth]{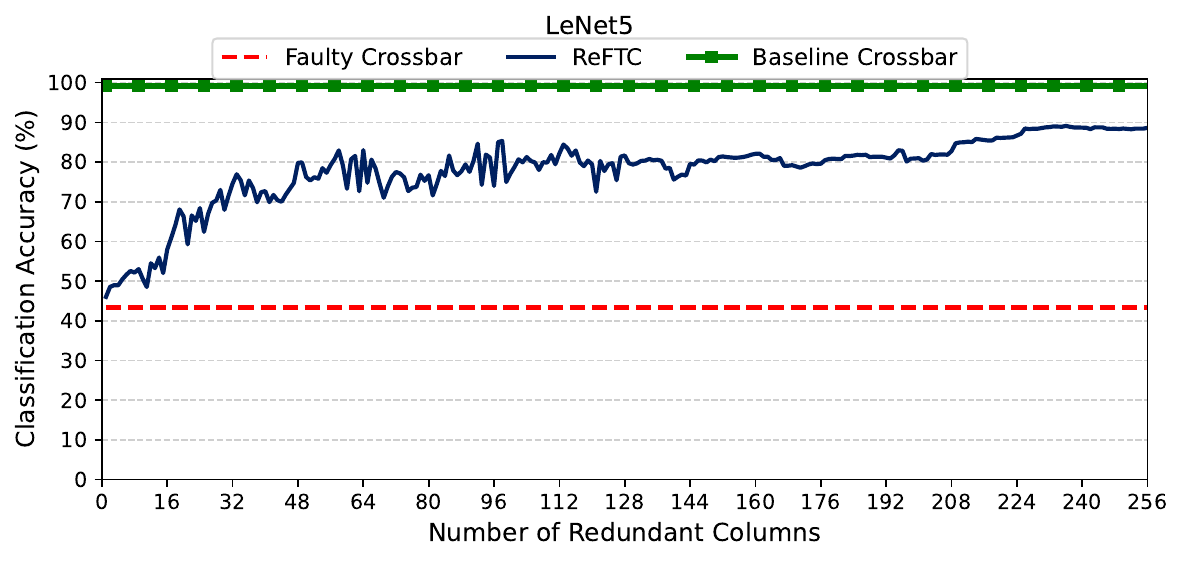}
}
	\caption{Classification accuracy and latency of ReFTC for the \textit{LeNet5} model with varying numbers of (a) redundant rows and (b) redundant columns.}
	\label{Fig12:multi2}
\end{figure*}

\begin{figure*}[!t]
	\centering
	\subfloat[\label{Fig13-a}]{
		\includegraphics[width=0.45\linewidth]{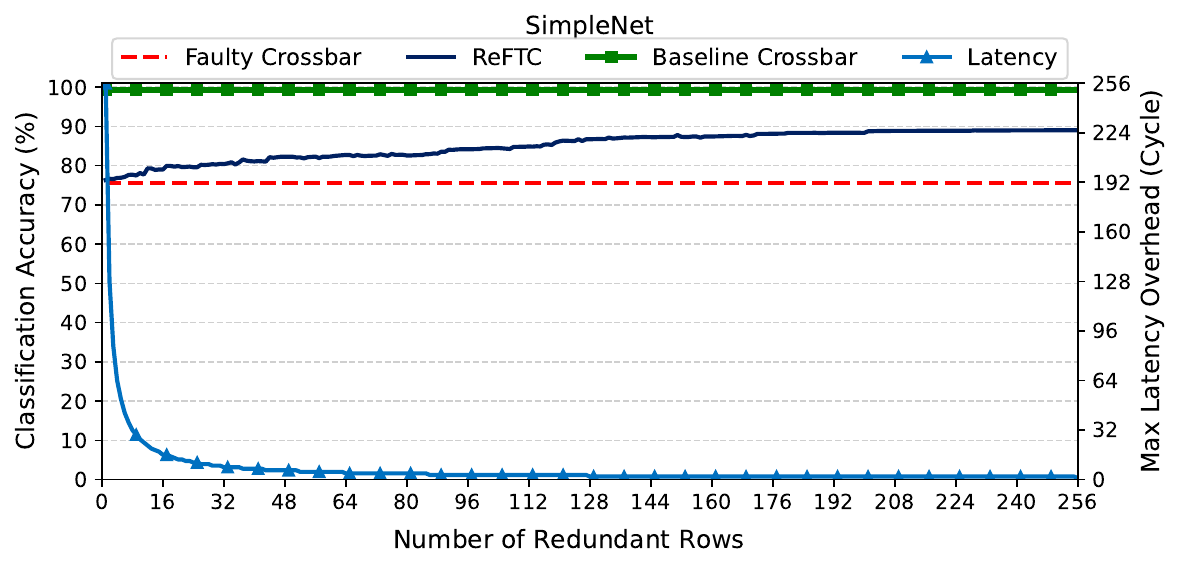}
	}
	\subfloat[\label{Fig13-b}]{
		\includegraphics[width=0.45\linewidth]{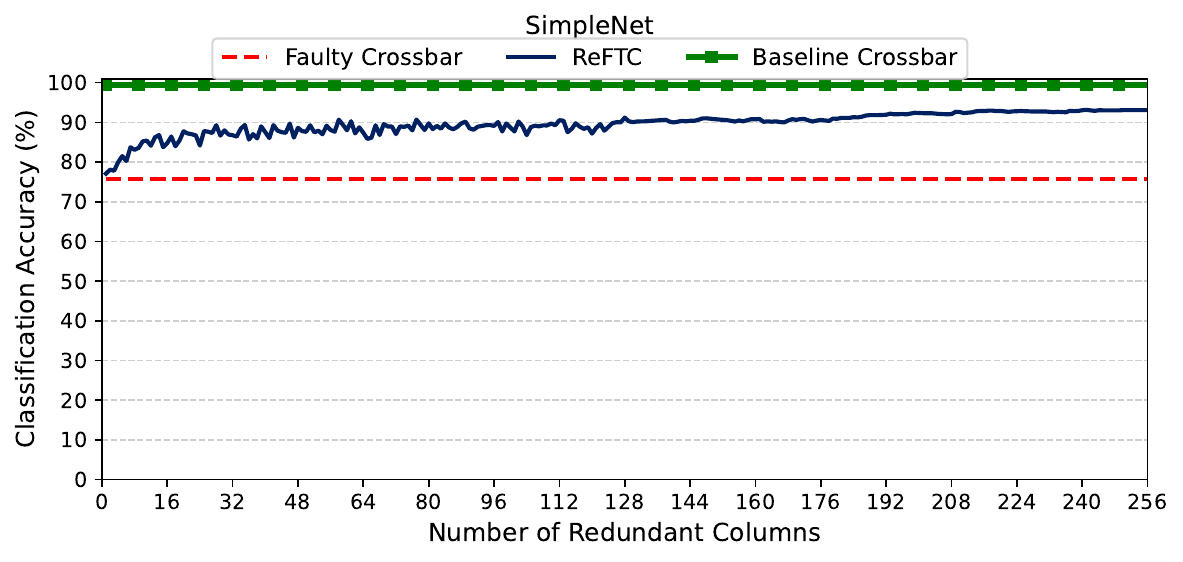}
	}
	\caption{Classification accuracy and latency of the ReFTC for the \textit{SimpleNet} model with varying numbers of (a) redundant rows and (b) redundant columns.}
	\label{Fig13:multi2}
\end{figure*}

To analyze the impact of the number of redundant rows and columns on the fault tolerance of the proposed ReFTC architecture, the learning parameters of two CNNs, \textit{LeNet5} and \textit{SimpleNet}, were mapped onto ReRAM crossbars. Each row and column can tolerate a certain number of faults without affecting the classification accuracy. However, if the number of faults exceeds a specific threshold, the redundant rows and columns are not able to tolerate all of faults. Therefore, the classification accuracy degrades as the number of faults increases. Accordingly, when the fault count is low, the classification accuracy does not change. A recent study reported that approximately 9.04\% and 1.75\% of ReRAM cells are affected by \textit{SaH} and \textit{SaL} faults, respectively~\cite{Ching-YiChen2015}. Therefore, \textit{SaL} and \textit{SaH} faults were randomly injected into 2\% and 10\% of the cells in a $256\times256$ crossbar with 65,536 cells, corresponding to 1310 and 6553 faulty cells, respectively. The classification accuracy of the \textit{LeNet5} and \textit{SimpleNet}, after being mapped onto a fault-free ReRAM accelerator, was 99.19\% and 99.36\%, respectively. However, after injecting faults into the ReRAM cells, the accuracy dropped to 43.34\% and 75.66\%, respectively. The proposed mapping and scheduling algorithms were executed for redundant rows and columns ranging from 0 to 256. The classification accuracy and maximum delay across 1000 experimental iterations of the proposed method for \textit{LeNet5} and \textit{SimpleNet}, with varying numbers of rows and columns, are presented in Figs.~\ref{Fig12:multi2} and~\ref{Fig13:multi2}, respectively. 

As shown in Figs.~\ref{Fig12-a} and~\ref{Fig13-a}, by increasing the number of redundant rows, the classification accuracy of \textit{LeNet5} and \textit{SimpleNet} increases gradually. The maximum accuracy increments for \textit{SimpleNet} and \textit{LeNet5} are about 13.36\% and 38.01\%, respectively. Moreover, the worst-case latency overhead of VMM operation decreases proportionally to the inverse of the number of rows. As shown in Figs.~\ref{Fig12-b} and~\ref{Fig13-b}, by increasing the number of redundant columns, the classification accuracy increases more than with redundant rows. Because, redundant columns can be utilized to tolerate both types of \textit{SaH} and \textit{SaL} faults. The maximum accuracy increment is about 17.49\% for \textit{SimpleNet} and 45.77\% for \textit{LeNet5}. Since redundant columns introduce no latency overhead, they are excluded from Figs.~\ref{Fig12-b} and~\ref{Fig13-b}. 

\begin{figure}[!t]
	\centering
	\subfloat[\label{Fig14-a}]{
		\includegraphics[width=0.4\linewidth]{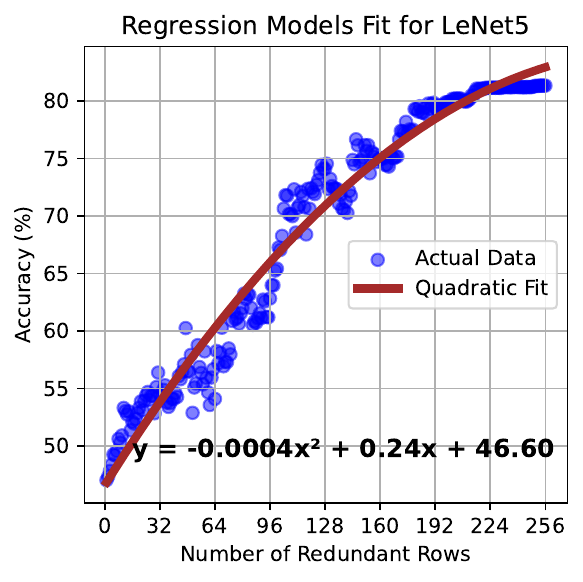}
	}
	\subfloat[\label{Fig14-b}]{
		\includegraphics[width=0.4\linewidth]{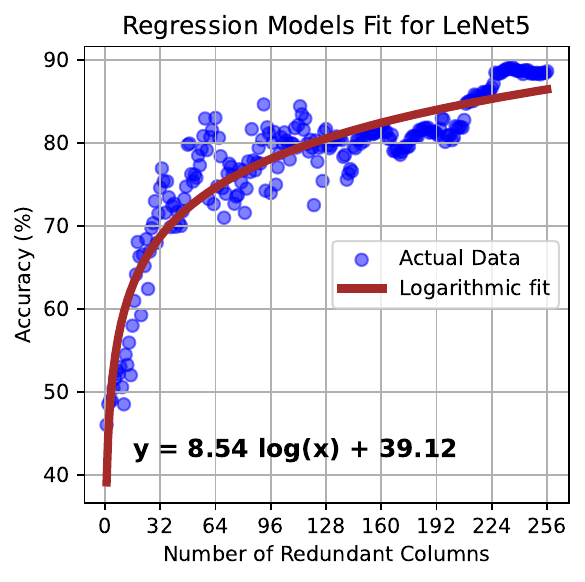}
	}
	
	\subfloat[\label{Fig14-c}]{
	\includegraphics[width=0.4\linewidth]{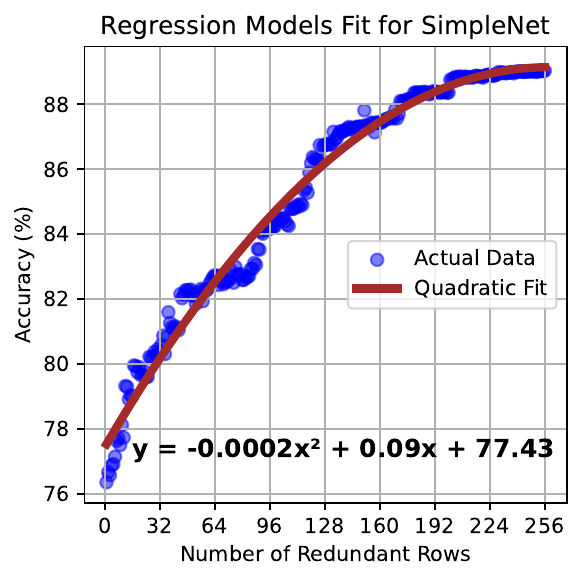}
	}
	\subfloat[\label{Fig14-d}]{
		\includegraphics[width=0.4\linewidth]{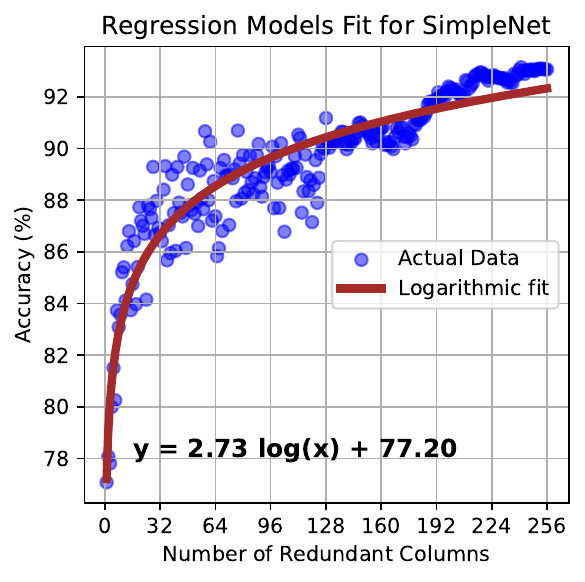}
	}
	
	\caption{Regression models for the relationship between the number of a) redundant rows in \textit{LeNet5}, b) redundant columns in \textit{LeNet5}, c) redundant rows in \textit{SimpleNet} and d) redundant columns in \textit{SimpleNet}.}
	\label{Fig14:multi4}
\end{figure}

Various linear and nonlinear regression models were examined to analyze the relationship between the number of rows/columns and classification accuracy and model $Accuracy_{row} (r)$ and $Accuracy_{col} (c)$. The appropriate regression model for the relationship between the number of rows and classification accuracy is a nonlinear quadratic regression, with coefficient of determination ($R^2$) values of 0.98 and 0.96 for the \textit{SimpleNet} and \textit{LeNet5} models, respectively, indicating that the model fits the data well. The appropriate regression model for the relationship between the number of columns and classification accuracy is a nonlinear logarithmic regression, with an $R^2$ value of 0.86 for both models. In Fig.~\ref{Fig14:multi4}, the quadratic and logarithmic regression models for the relationship between the number of redundant rows and columns in the \textit{SimpleNet} and \textit{LeNet5} models are shown. The fault-tolerant accuracy model $Accuracy_{FT} (r,c)$, defined in Eq.~\ref{Eq20}, was used to estimate the classification accuracy of \textit{SimpleNet} and \textit{LeNet5} running on the ReFTC architecture. The average estimation error was about 5.23\% for \textit{SimpleNet} and 4.78\% for \textit{LeNet5} compared to simulation result of 100 random configurations of $(r,c)$.

\subsection{Evaluation of Fault-Aware Weight Mapping in Dual ReRAM Crossbar (FADM)}
To evaluate the efficiency of the proposed FADM method in a dual positive and negative crossbar, varying number of \textit{SaH} and \textit{SaL} faults are injected into positive and negative ReRAM crossbars. Then, the fault-aware conductance values for faulty cells are calculated based on Eqs.~\ref{Eq26} to~\ref{Eq29} and remapped to their corresponding dual cells.

Variations and faults in memory cells can cause discrepancies between the effective conductance value and the ideal conductance we want to program. To measure the deviation in weight storage and programming, the summed weight variations ($SWV$) metric is employed. $SWV$ quantifies the difference between the ideal weight value ($W$) and the effective weight value ($W^{eff}$)~\cite{Ma2023}. It is computed by,
\begin{equation}
	\text{SWV} = \sum_{i=1}^{N} \sum_{j=1}^{M} \left| w_{i,j}^{\text{eff}} - w_{i,j} \right|
	\label{Eq30}
\end{equation}
where $w_{(i,j)}$ represents the weight value at position $(i,j)$ in the weight matrix $W$. A higher $SWV$ value indicates a greater deviation from the ideal weights, leading to increased errors and, consequently, a more significant accuracy loss.

\begin{figure}[!t]
	\centering
	\includegraphics[width=0.45\textwidth]{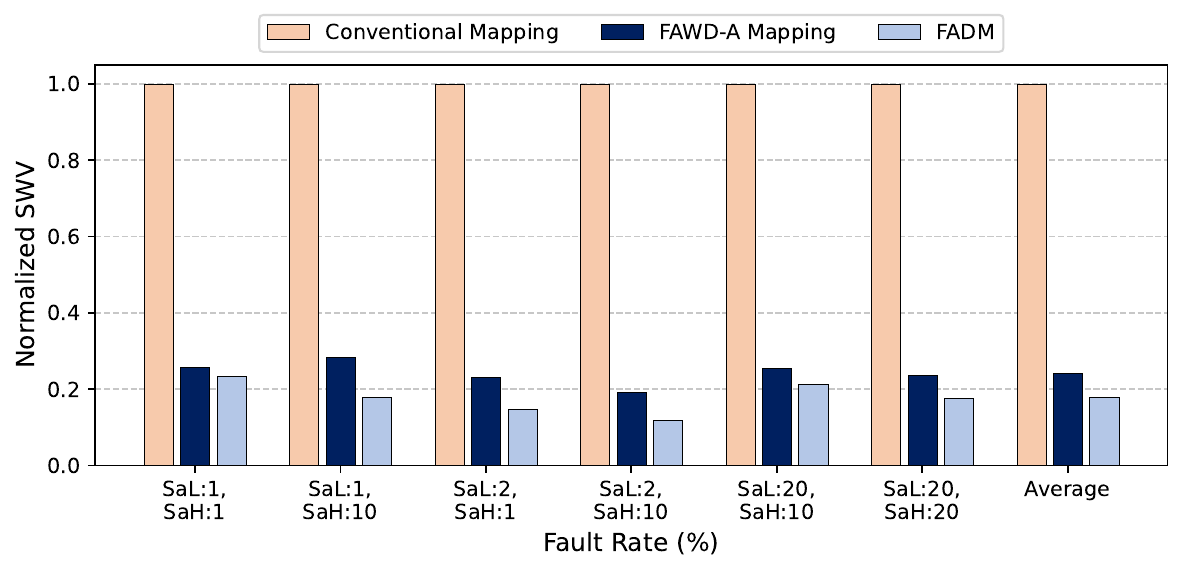}
	\caption{The result of SWV in different fault rates in a $256\times256$ ReRAM crossbar.}
	\label{Fig15-SWF-FADM}
\end{figure}

The results of 20,000 fault injection iterations are presented in Fig.~\ref{Fig15-SWF-FADM}, covering various fault percentages. These experiments involve randomly assigned weight matrix values and the injection of \textit{SaH} and \textit{SaL} faults at random locations. The results of normalized $SWV$ for conventional mapping method, Fault-Aware Weight Decomposition for analog ReRAM arrays (FAWD-A) mapping method~\cite{Bang2023}, and our proposed FADM are reported in Fig.~\ref{Fig15-SWF-FADM}. On average, our proposed FADM can further reduce $SWV$ by 82\% (5 times) compared to conventional mapping method, while FAWD-A reduce SWV by 75\%.

Additionally, we evaluated the fault coverage capability of FADM under various fault rates by measuring the ratio of tolerable and intolerable fault cases in the dual-crossbar structure, as reported in Table~\ref{Table3:tol-vs-intol}. The results show that although FADM is constrained by the weight sign, fault type, and fault distribution, it can still effectively tolerate approximately 50\%–60\% of all fault cases.
At low fault rates (e.g., \textit{SaH}: 1\%, \textit{SaL}: 1\%), most non-ideal cases correspond to single faults. Since the distributions of positive and negative weights in the neural network are approximately balanced, a single fault has nearly equal probabilities of resulting in either a tolerable state or an intolerable state. As a result, the tolerable ratio remains close to 50\%, which is consistent with the reported value of 50.35\% in Table~\ref{Table3:tol-vs-intol}.
As the fault rates increase (e.g., 20\%), the probability of simultaneous double faults increases significantly. In the differential dual-crossbar structure, symmetric double faults such as (\textit{SaH}, \textit{SaH}) and (\textit{SaL}, \textit{SaL}) naturally compensate for each other, resulting in an effective differential conductance close to zero. Since a large portion of the weights in trained CNNs are either exactly zero or close to zero, these symmetric double-fault conditions become functionally aligned with the intended weight values and are therefore classified as ideal or effectively tolerable states. Consequently, although the absolute number of faulty cells increases, the proportion of tolerable states also rises, reaching 60.96\% at the 20\% fault rate.
These results demonstrate that FADM can maintain substantial fault coverage even under severe fault conditions, making it a practical complementary technique that can be effectively integrated with other fault-tolerant methods to further improve system reliability.

\begin{table}[!t]
	\centering
	\caption{The percentages of tolerable vs intolerable fault cases of dual crossbar.}
	\begin{adjustbox}{max width=1\linewidth}
		\begin{tabular}{
				>{\raggedright\arraybackslash}m{2.5cm} 
				>{\centering\arraybackslash}m{1cm} 
				>{\centering\arraybackslash}m{1cm}
				>{\centering\arraybackslash}m{1cm}
				>{\centering\arraybackslash}m{1cm}
				>{\centering\arraybackslash}m{1cm}
				>{\centering\arraybackslash}m{1cm}    
			}
			\toprule
			Fault Rates (\%) & SaH:1, SaL:1 & SaH:1, SaL:2 & SaH:10, SaL:1 & SaH:10, SaL:2 & SaH:10, SaL:20 & SaH:20, SaL:20 \\
			\midrule
			Tolerable States & 50.35\% & 50.78\% & 52.85\% & 53.18\% & 58.18\% & 60.96\%\\
			Intolerable States   & 49.64\% & 49.22\% & 47.15\%  & 46.82\% & 41.82\% & 39.04\%\\
			\bottomrule
		\end{tabular}
	\end{adjustbox}
	\label{Table3:tol-vs-intol}
\end{table}

\begin{figure}[!t]
	\centering
	\includegraphics[width=0.45\textwidth]{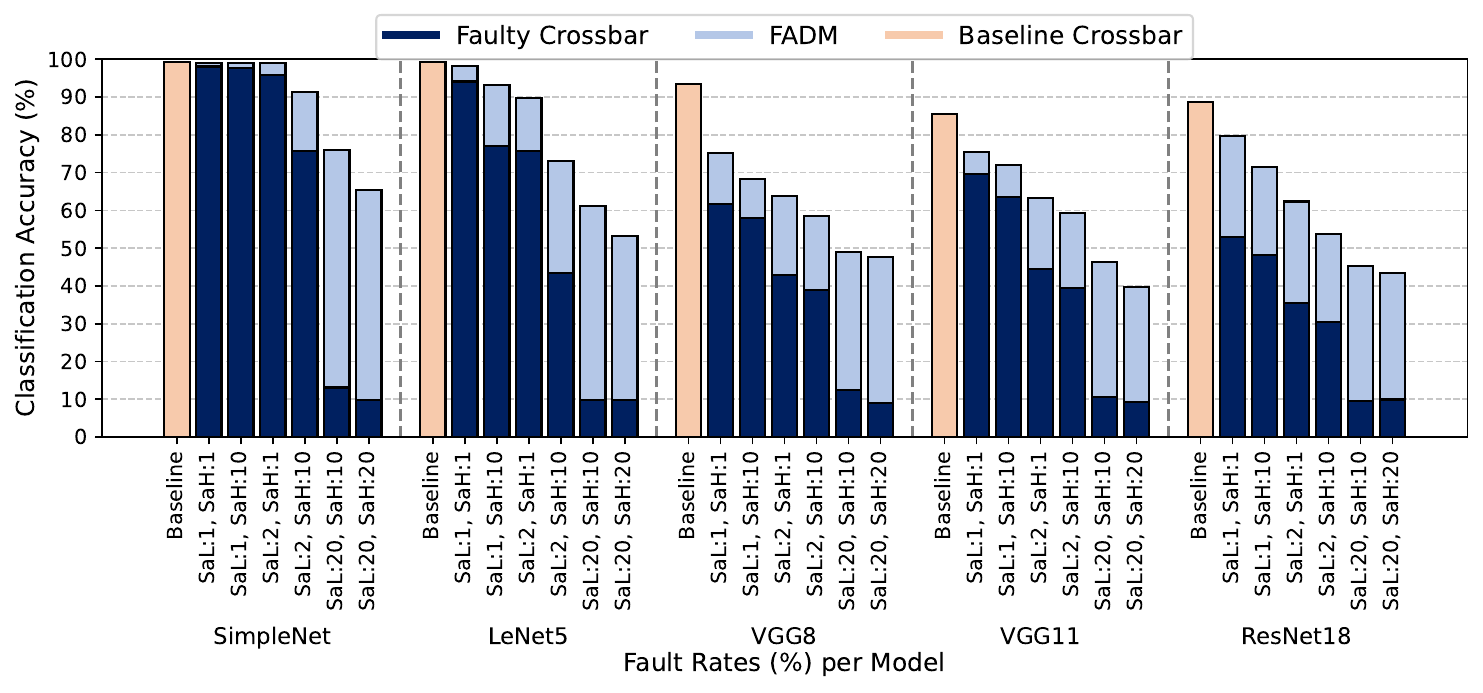}
	\caption{Classification Accuracy of FADM mapping to a $256\times256$ main crossbar for different fault rate per model.}
	\label{Fig16-FADM}
\end{figure}

To evaluate the proposed FADM on classification accuracy of CNNs, varying numbers of \textit{SaH} and \textit{SaL} faults were injected into the ReRAM crossbar. The accuracy of the non-faulty crossbar, faulty accuracy, and accuracy incremented after applying the proposed method are shown Fig.~\ref{Fig16-FADM}. As the number of faults increases, the classification accuracy decreases; however, the proposed method effectively mitigates the impact of \textit{SaH} and \textit{SaL} faults. On average, FADM increases the accuracy of \textit{SimpleNet} and \textit{LeNet5} by approximately 17.04\% and 19.9\%, and for \textit{VGG8}, \textit{VGG11}, and \textit{ResNet18}, the accuracy increases by about 17.48\%, 14.91\%, and 21.16\%, respectively.
 

\begin{figure*}[!t]
	\centering
	\subfloat[\label{Fig17-a}]{
		\includegraphics[width=0.45\linewidth]{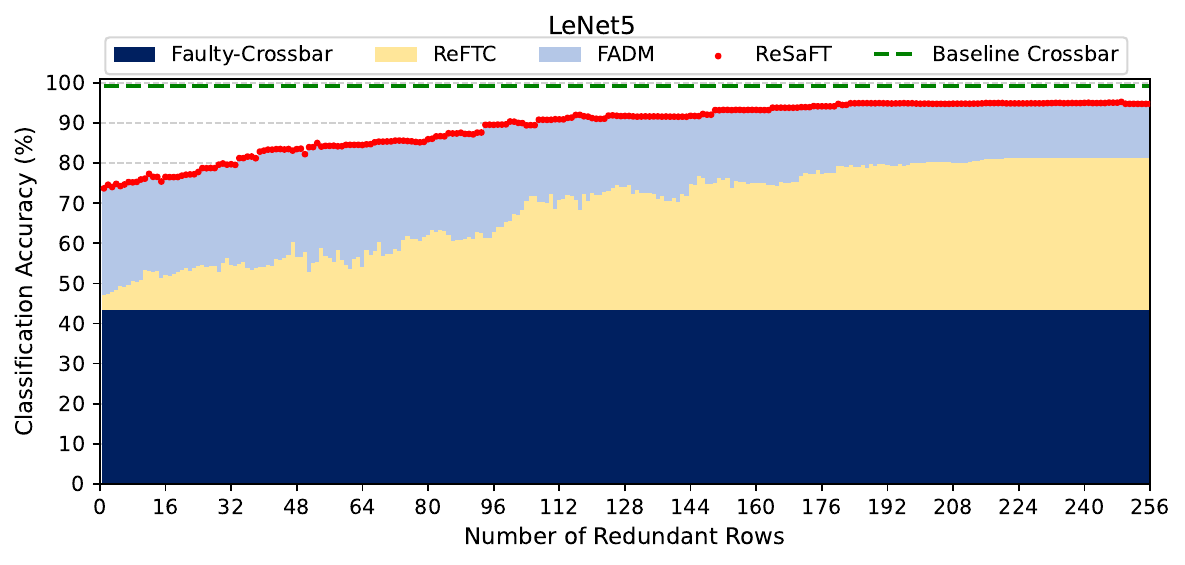}
	}
	\subfloat[\label{Fig17-b}]{
		\includegraphics[width=0.45\linewidth]{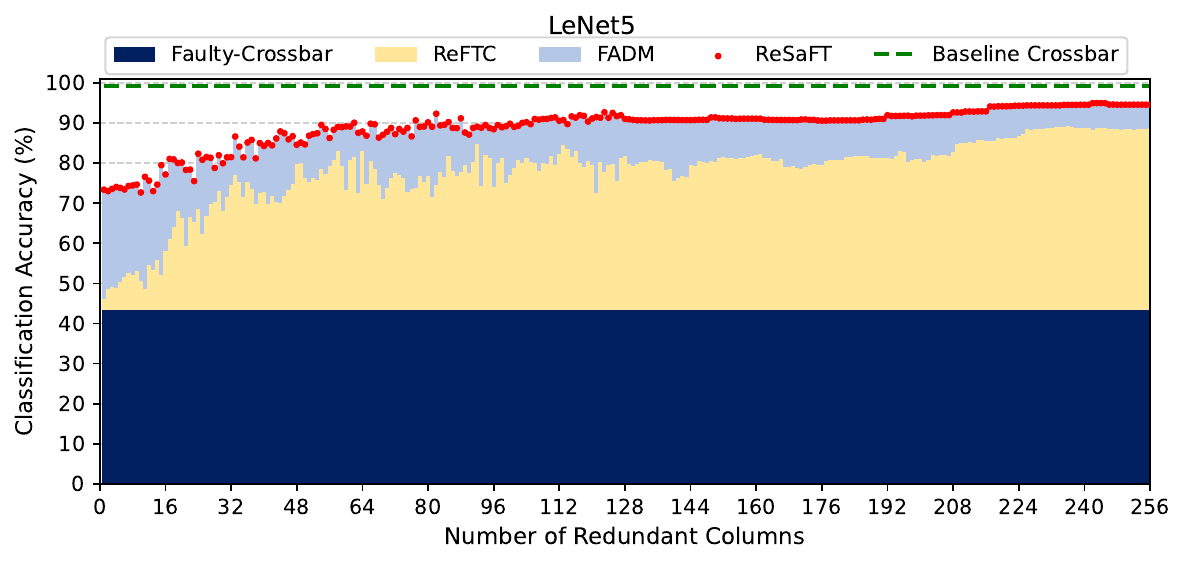}
	}
	\caption{Classification accuracy of FADM applied to the ReFTC for the \textit{LeNet5} model with varying numbers of (a) redundant rows and (b) redundant columns.}
	\label{Fig17:multi2}
\end{figure*}

\begin{figure*}[!t]
	\centering
	\subfloat[\label{Fig18-a}]{
		\includegraphics[width=0.45\linewidth]{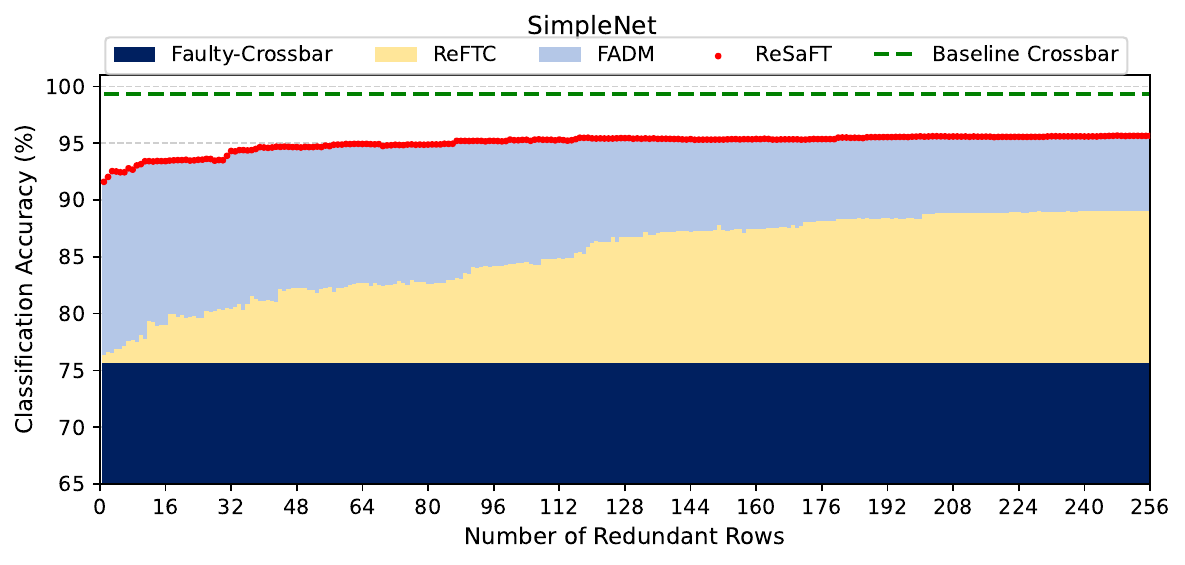}
	}
	\subfloat[\label{Fig18-b}]{
		\includegraphics[width=0.45\linewidth]{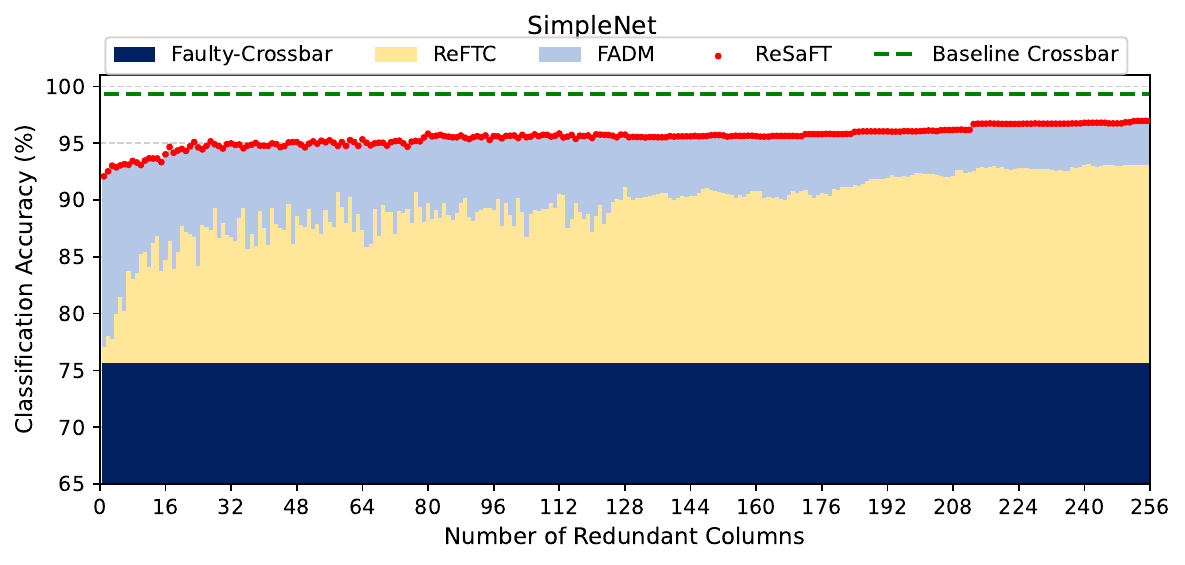}
	}
	\caption{Classification accuracy of FADM applied to the ReFTC for the \textit{SimpleNet} model with varying numbers of (a) redundant rows and (b) redundant columns.}
	\label{Fig18:multi2}
\end{figure*}

In the next step, the proposed FADM method is applied to the redundancy-based fault-tolerant ReRAM crossbar, referred to as the ReSAFT. The results of classification accuracy for \textit{LeNet5} and \textit{SimpleNet} after FADM with varying numbers of redundant rows and columns are shown in Figs.~\ref{Fig17:multi2} and~\ref{Fig18:multi2}. The average, minimum, and maximum percentages of accuracy increment based on the faulty crossbar, as well as the accuracy of the multi-FT-crossbar, are reported in Table~\ref{Table3:accuracy}.

As shown in Figs.~\ref{Fig17-a} and~\ref{Fig18-a}) and reported in Table~\ref{Table3:accuracy}, applying the FADM mapping in the redundancy-based FT-crossbar has significantly increased classification accuracy. Even in cases where a small number of rows or columns could not increase the accuracy, the FADM mapping still increased the accuracy to a greater extent. The average classification accuracy for different numbers of redundant rows and columns in the multi-FT-crossbar is 89.17\% and 89.02\% for \textit{LeNet5}, and 95.02\% and 95.55\% for \textit{SimpleNet}, respectively.

\begin{table}[t]
	\centering
	\large
	\caption{Classification accuracy results of the FADM mapping applied to the ReFTC.}
	\begin{adjustbox}{max width=1\linewidth}
		\begin{tabular}{
				>{\raggedright\arraybackslash}m{3cm} 
				>{\raggedright\arraybackslash}m{1cm} 
				>{\centering\arraybackslash}m{2cm}  
				>{\centering\arraybackslash}m{2cm}  
				>{\centering\arraybackslash}m{2cm}  
				>{\centering\arraybackslash}m{2cm}  
			}
			\toprule
			& & \multicolumn{2}{c}{LeNet5} & \multicolumn{2}{c}{SimpleNet} \\
			\cmidrule(lr){3-4} \cmidrule(lr){5-6}
			& & Redundant Row & Redundant Column & Redundant Row & Redundant Column \\
			\midrule
			\multicolumn{2}{>{\raggedright\arraybackslash}m{5cm}}{Baseline Crossbar Accuracy} 
			& 99.19\% & 99.19\% & 99.36\% & 99.36\% \\
			\midrule
			
			\multicolumn{2}{>{\raggedright\arraybackslash}m{5cm}}{Faulty Crossbar Accuracy}
			& 43.34\% & 43.34\% & 75.66\% & 75.66\% \\
			\midrule
			
			\multirow{3}{=}{ReFTC Accuracy Increment} 
			& Average & 25.59\% & 34.69\% & 9.58\% & 13.98\% \\
			& Min & 3.69\% & 2.69\% & 0.69\% & 1.41\% \\
			& Max & 38.01\% & 45.77\% & 13.37\% & 17.49\% \\
			\midrule
			
			\multirow{3}{=}{FADM Accuracy Increment} 
			& Average & 20.24\% & 10.98\% & 9.78\% & 5.91\% \\
			& Min & 13.37\% & 4.41\% & 6.57\% & 3.52\% \\
			& Max & 31.05\% & 28.04\% & 15.97\% & 15.22\% \\
			\midrule
			
			\multirow{3}{=}{ReSAFT Accuracy} 
			& Average & 89.17\% & 89.02\% & 95.02\% & 95.55\% \\
			& Min & 73.67\% & 72.65\% & 91.60\% & 92.08\% \\
			& Max & 95.23\% & 94.92\% & 95.67\% & 96.96\% \\
			\bottomrule
		\end{tabular}
	\end{adjustbox}
	\label{Table3:accuracy}
\end{table}

\subsection{MOO Problem Analysis}
To solve the proposed MOO problem, the NSGA-II is utilized, which is a widely used evolutionary algorithm for multi-objective optimization. This algorithm efficiently explores the design space and identifies a set of optimal solutions, known as the Pareto front, for the number of redundant rows and columns, ensuring an effective trade-off among all objectives.

\begin{figure}[!t]
	\centering
	\includegraphics[width=0.3\textwidth]{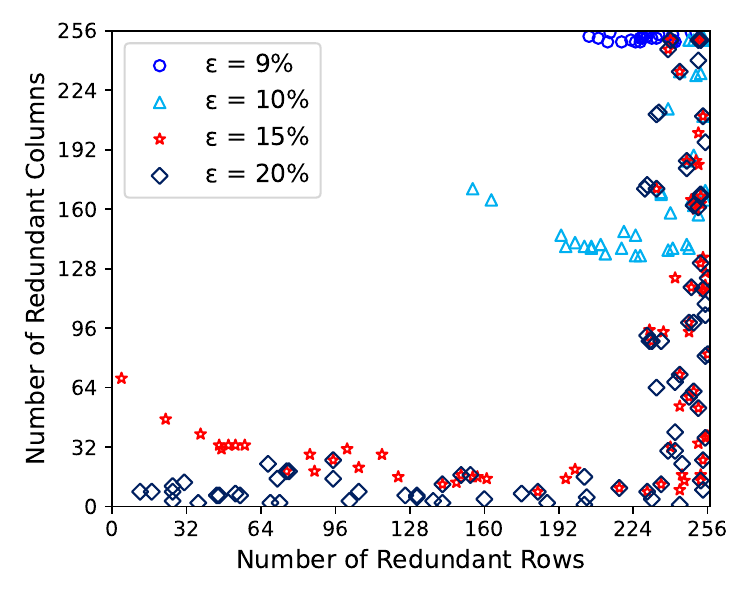}
	\caption{Semi-optimal combinations of redundant rows and columns for \textit{SimpleNet}.}
	\label{Fig19-pareto-front}
\end{figure}

The result of optimal number of rows and columns for different value of $\varepsilon$ on \textit{SimpleNet} model, based on NSGA-II algorithm, is shown in Fig.~\ref{Fig19-pareto-front}. As mentioned above, there is no best solution but a set of semi-optimal solutions are provided by the algorithm that cover as much as the objective space as possible. For example, a $256\times256$ main crossbar, and the constraint of $r,c\in[0,256]$, the design space contains 66,049 possible solutions that only limited number of them satisfy the objective functions. The number of semi-optimal solutions shown in Table~\ref{Table4:pareto-front}. Since $\varepsilon$ is a constraint in the problem, smaller values of $\varepsilon$ make the problem more challenging, leading to fewer optimal solutions. Conversely, as $\varepsilon$ increases, the number of solutions grows, and they better fulfill the problem's objectives. Therefore, the designer should select a solution from the Pareto front based on the specific application requirements and the relative importance of each objective.

\begin{table}[!t]
	\centering
	\caption{The number of optimal solutions for redundant rows and columns for \textit{SimpleNet} and \textit{LeNet5}.}
	\begin{adjustbox}{max width=1\linewidth}
		\begin{tabular}{
			>{\raggedright\arraybackslash}m{2.5cm} 
			>{\centering\arraybackslash}m{2cm} 
			>{\centering\arraybackslash}m{2cm}  
			>{\centering\arraybackslash}m{2cm}  
			>{\centering\arraybackslash}m{2cm}    
			}
			\toprule
			Model & $\varepsilon = 9\%$ & $\varepsilon = 10\%$ & $\varepsilon = 15\%$ & $\varepsilon = 20\%$ \\
			\midrule
			SimpleNet & 23 & 41 & 73 & 88 \\
			LeNet5    & 18 & 38 & 79 & 91 \\
			\bottomrule
		\end{tabular}
	\end{adjustbox}
	\label{Table4:pareto-front}
\end{table}

\begin{figure}[!t]
	\centering
	\subfloat[\label{Fig20-a}]{
		\includegraphics[width=0.45\linewidth]{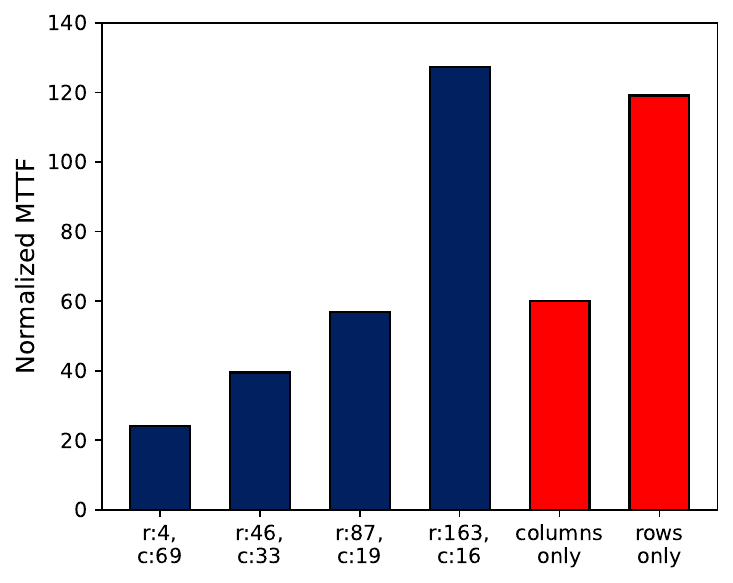}
	}
	\subfloat[\label{Fig20-b}]{
		\includegraphics[width=0.45\linewidth]{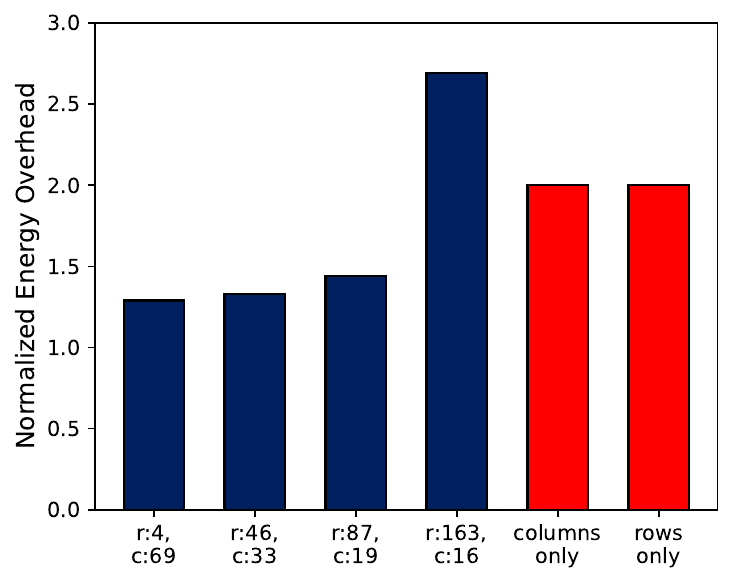}
	}
	
	\subfloat[\label{Fig20-c}]{
		\includegraphics[width=0.45\linewidth]{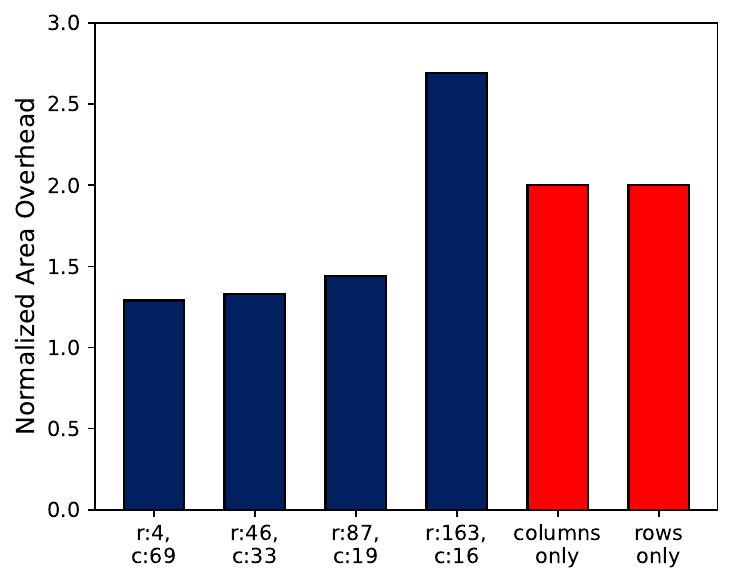}
	}
	\subfloat[\label{Fig20-d}]{
		\includegraphics[width=0.45\linewidth]{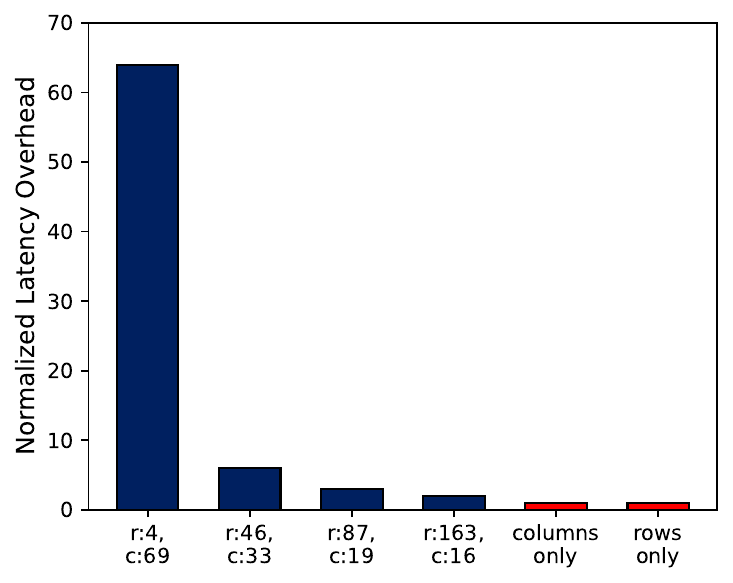}
	}
	
	\caption{Normalized (a) MTTF improvement, (b) energy overhead, (c) area overhead, and (d) latency overhead for four optimized ReFTC configurations compared to the baseline ReRAM crossbar under $\varepsilon=15\%$.}
	\label{Fig20:multi4}
\end{figure}

We compared the proposed method with two representative baseline configurations with just 256 redundant rows (rows-only) or 256 redundant columns (columns-only), same as the size of main crossbar. These configurations represent the fundamental row-wise and column-wise redundancy strategies and serve as fair baselines for evaluation. Fig.~\ref{Fig20:multi4} illustrates the normalized results of MTTF improvement and energy consumption, area, and latency overhead for four different optimized ReFTC configurations, compared to two baseline configurations.

The results are normalized to an ideal fault-free ReRAM without any redundancy. We assume that the accuracy constraint is $\varepsilon=15\%$. Each configuration offers a distinct trade-off between reliability and overheads, allowing designers to select the most suitable option based on the system requirements. For instance, using 87 redundant rows and 19 redundant columns yields a $56\times$ improvement in MTTF with a 43\% increase in energy consumption compare to the baseline. The area and energy overheads are reduced by approximately 28\% compared to rows-only or columns-only configurations, while the latency overhead increases by about $3\times$. In contrast, a configuration with 163 redundant rows and 16 redundant columns achieves a higher MTTF improvement of $127\times$, at the cost of a 169\% increase in energy consumption. Since redundant columns mask both \textit{SaH} and \textit{SaL} faults, columns-only configuration achieves higher MTTF, but has more area and energy overhead. 

Table~\ref{Table5:accuracy_resaft} presents the classification accuracy of the \textit{SimpleNet} model for four optimized ReFTC configurations with FADM. While faults reduce the baseline accuracy from 99.33\% to 75.66\%, by applying the proposed ReFTC, the accuracy improves by an average of 18.71\%. Furthermore, incorporating the FADM algorithm yields an additional average improvement of 2.98\%, resulting in a total accuracy enhancement of 21.7\% achieved by the proposed ReSAFT scheme.
\begin{table}[!t]
	\centering
	\caption{Classification accuracy for optimized ReFTC configurations and two baseline model with FADM on the \textit{SimpleNet} model.}
	\label{Table5:accuracy_resaft}
	\begin{adjustbox}{max width=1\linewidth}
		\begin{tabular}{
				>{\raggedright\arraybackslash}m{2cm} 
				>{\centering\arraybackslash}m{2cm} 
				>{\centering\arraybackslash}m{2.5cm}  
				>{\centering\arraybackslash}m{2cm}  
				>{\centering\arraybackslash}m{2cm}    
			}
			\toprule
			  & Baseline & Faulty Crossbar & ReFTC & ReSAFT \\
			\midrule
			r:4, c:69     & 99.33\% & 75.66\% & 93.59\% & 96.82\% \\
			r:46, c:33    & 99.33\% & 75.66\% & 93.84\% & 97.23\% \\
			r:87, c:19    & 99.33\% & 75.66\% & 94.33\% & 97.69\% \\
			r:163, c:165  & 99.33\% & 75.66\% & 95.72\% & 97.86\% \\
			\rowcolor{red!15}
			columns-only         & 99.33\% & 75.66\% & 91.07\% & 95.96\% \\
			\rowcolor{red!15}
			rows-only         & 99.33\% & 75.66\% & 88.03\% & 94.65\% \\
			\bottomrule
			\multicolumn{5}{l}{\textbf{r}: number of redundant rows, \textbf{c}: number of redundant columns}
		\end{tabular}
\end{adjustbox}	
\end{table}

\section{Conclusion}
\label{Concl}
This work addresses a critical reliability bottleneck in analog ReRAM-based PIM accelerators for CNN inference by targeting SAFs. We present fault-tolerant ReRAM crossbar architecture that combines hardware redundancy, through both redundant rows and columns, with adaptive mapping and input voltage scheduling methods. This architecture effectively mitigates both \textit{SaH} and \textit{SaL} faults with minimal impact on design overhead and without sacrificing inference accuracy. We conducted a comprehensive evaluation to investigate how varying the number of redundant rows and columns affects key design metrics of ReRAM Accelerator, such as MTTF, energy consumption, latency, area, and the classification accuracy of CNNs. Based on the modeling of these metrics, an optimization problem with NSGA-II algorithm was presented to find the optimal number of redundant rows and columns that cover as much as the objective space as possible. Additionally, a fault-aware mapping and input scheduling to redundant rows and columns was performed. Furthermore, the incorporation of a fault-aware weight mapping within both positive and negative crossbars ensures further compensating for the accuracy degradation caused by SAFs. Based on the Pareto front derived from the optimization problem, four distinct configurations that fulfill the problem's objectives, were analyzed, resulting in an improvement of approximately 22.39\% in the inference accuracy of the \textit{SimpleNet} model. Overall, our proposed ReSAFT scheme substantially improves the reliability and computational efficiency of analog ReRAM-based accelerators, making them more suitable for deployment in safety-critical and high-precision deep learning applications.

\bibliographystyle{elsarticle-num} 
\bibliography{mybib_abbr}

\end{document}